\documentclass[preprint,12pt,eqsecnum,nofootinbib,amsmath,amssymb]{revtex4}

\newcommand{\datechange}{10/21/2021}
\newcommand{\preprintnumber}{}

\newcommand{\mytitle}{Grover's Algorithm with Diffusion and Amplitude Steering}

\usepackage{graphicx}  
\usepackage{dcolumn}  
\usepackage{bm}             
\usepackage{latexsym}  
\usepackage{fancyhdr}  
\usepackage{wrapfig}
\usepackage{comment}
\usepackage{dsfont}
\usepackage{mathtools}
\usepackage{datetime}
\usepackage{mathrsfs}
\usepackage{amsbsy}
\usepackage{amsmath}
\usepackage{amssymb}
\usepackage{bbold}

\newcommand{\smA}{{\scriptscriptstyle \rm A}}
\newcommand{\smB}{{\rm\scriptscriptstyle B}}

\newcommand{\smT}{{\rm\scriptscriptstyle T}}

\newcommand{\bodyskip}{\baselineskip 18pt plus 1pt minus 1pt}
\newcommand{\bibskip}{\baselineskip16pt plus 1pt minus 1pt}
\newcommand{\tableofcontentsskip}{\baselineskip 14pt plus 1pt minus 1pt}
\newcommand{\footnoteskip}{\baselineskip 12pt plus 1pt minus 1pt}
\newcommand{\abstractskip}{\baselineskip 13pt plus 1pt minus 1pt}
\newcommand{\titleskip}{\baselineskip 18pt plus 1pt minus 1pt}
\newcommand{\affiliationskip}{\baselineskip 15pt plus 1pt minus 1pt}

\begin{document}

\baselineskip 20pt plus 1pt minus 1pt

 \title{\titleskip
   \mytitle
 }

 \author{Robert L Singleton Jr$^\dagger$,  Michael L Rogers,  and David L Ostby}

 \affiliation{\affiliationskip
   SavantX Research Center\\
  Santa Fe,   NM  87501\\
  $\dagger$\,{\rm robert.singleton@savantx.com}
 }

 \date{\datechange}

 \begin{abstract}
 \abstractskip
 \vskip0.3cm 

\noindent
We review the basic theoretical underpinnings of the Grover 
algorithm,  providing  a rigorous and well motivated derivation.  
We then present a generalization of Grover's algorithm that 
searches an arbitrary subspace of the multi-dimensional Hilbert 
space using a diffusion operation and an amplitude amplification 
procedure that has been biased by unitary {\em steering 
operators}.  We also outline a generalized Grover's algorithm 
that takes into account higher level correlations that could exist 
between database elements.  In the traditional Grover algorithm,  
the Hadamard gate selects a uniform sample of computational 
basis elements when performing the phase selection and diffusion.   
In contrast,   steered operators bias the selection process,  thereby 
providing  more flexibility in selecting the target state.   Our method 
is a generalization of the recently proposal pattern matching 
algorithm of Hiroyuki {\em et al.}.

 \end{abstract}

\maketitle

\pagebreak
\tableofcontentsskip
\tableofcontents

\newpage
\bodyskip

\pagebreak
\clearpage

\section{Introduction}

Searching an unstructured data base of $N$ elements by classical 
means takes an average of $N/2$ calls to the database. In contrast,  
{\em Grover's algorithm}\,\cite{Grover} is a quantum search algorithm 
that requires of order $\sqrt{N}$ calls.  While this is only a quadratic 
improvement,  for large values of $N$ this can be a substantial savings.   
For example,  a data base with \hbox{$N=10,000$} elements could 
be searched in just $100$ quantum calls,  compared 
to an average of 5,000 classical calls.  The essential ingredients
of Grover's algorithm are {\em amplitude selection} and the 
subsequent {\em diffusion} of an initial or trial quantum state.  
Together,   these steps constitute a process known as {\em 
amplitude amplification},  first introduced in Refs.~\cite{gilles,  
grover_amp,  gilles_two}.   In this section, we review the basic 
theoretical underpinnings of the Grover algorithm and amplitude 
amplification,   providing physically motivated derivations,
rather than the traditional proof-theorem construct of computer 
science and mathematics.   We generalize both the method 
and formalism of the amplitude amplification procedure. 
We go on to examine a number of variants  of the Grover 
algorithm, 
and  we recast amplitude amplification in such a way as to 
extract an arbitrary subspace of the $N$-dimensional Hilbert 
space.  We do this by biasing the diffusion operation and phase 
selection mechanism by unitary {\em steering} operators.   In 
the traditional Grover algorithm,   the Hadamard gate selects 
a uniform sample of computational basis elements when 
performing phase selection and diffusion.   In contrast,  the 
steering procedure biases these processes with a well chosen 
unitary operator,   thereby providing more flexibility for the 
algorithm.   We also apply our general formalism to give a 
concise derivation of the quantum pattern matching algorithm 
of Hiroyuki {\em et al.}\,\cite{image_grover}.  

We start our analysis with an overview of Grover's algorithm,
which consists of the following basic operations: (i) a trial wave 
function is selected and then used as a first guess,  (ii) the phase 
of the wave function in the direction of the target state is inverted,   
thereby imprinting the target state on the trial wave function,  
(iii) the resulting state is sent through a diffusion operation 
to enhance the marked component,  and (iv) the process is 
repeated until the target state is achieved with close to unit
probability.   In the traditional  Grover algorithm,  the target state 
is taken to be one of the computational basis states,  and the 
trial wave function is taken to be the Hadamard state consisting 
of a uniform superposition of basis elements.  While Grover's 
algorithm allows one to select a specific quantum state from 
the multi-qubit Hilbert space,   it has proven difficult to translate 
the algorithm into a realistic search engine.  Interestingly,  the 
difficulty of implementing a practical algorithm has led to 
numerous applications in other fields,  such as cryptography 
and signal processing,  but we would like to explore the possibility 
of using Grover's algorithm for its original intended purpose 
of a database search.  One of the problems with Grover search 
involves the difficulty of creating a quantum dictionary that 
maps the database entries into appropriate quantum states.  
This is a kind of chicken-and-egg problem:  if we knew the 
basis state corresponding to a specific database entry (and 
this basis state is precisely the information that Grover's 
algorithm returns),  then we would in fact already know the 
database entry,  and there would be no need for Grover's 
algorithm.  We shall call this the {\em dictionary problem}.  
The origin of this problem is that the traditional algorithm 
has only one preferred state,  namely the target state,  as the 
algorithm is initialized with a uniform superposition of basis 
states,  none of which are preferred.  If we were to weight 
these basis states in some preferred direction,  then we 
could avert the dictionary problem.   We do this by a 
mechanism we call {\em steering}.  Instead of employing  
a uniform sum over the basis states in selecting the initial 
guess,   we use a non-uniform biased sum determined 
by a well chosen unitary operator.  This biases the state 
selection,  thereby avoiding the chicken-and-egg problem,  
while simultaneously improving the efficiency of the algorithm.  

There are other problems in implementing a practical
Grover search.  In particular,  a database contains an 
exponential number of gates,  and a direct implementation
of the database would therefore  degrade the quantum 
advantage of the algorithm.  To overcome this {\em exponential 
gate} problem,  Ref.~\cite{aae21} introduces a new method
called {\em approximate amplitude encoding} (AAE).
This method approximates the exponentially large 
database by a constant depth  {\em parameterized 
quantum circuit} containing only a polygonal number 
of gates,  and the parameters of the circuit are trained 
using machine learning.  We will discuss this problem
throughout this paper,   particularly in the last section,  
where we reproduce the quantum circuit of Hiroyuki 
{\em et al.} from our formalism.  

This paper is organized as follows.  In Section~\ref{sec_notation} 
we introduce the general notion of steering operators.  In
Section~\ref{sec_grover_review} we derive the traditional
Grover's algorithm,  and in Section~\ref{sec_grover_steering} 
we generalize the algorithm to include what we call steering 
operators  meant to bias the diffusion and amplitude selection
processes.   Section~\ref{sec_grover_general} further generalizes 
Grover's algorithm to accommodate an arbitrary target set,
rather than a target consisting of a single computational basis 
element.  We also generalize the algorithm with a non-separable
kernel to account for possible higher order correlations between 
the search elements,  thereby leading to non-planar Grover
algorithms with speedups that are potentially better than 
quadratic.  In Section~\ref{sec_circuit} we construct several
quantum circuits that implement the generalized algorithms,
and in Section~\ref{sec_pattern} we derive the results of the 
pattern matching algorithm of Hiroyuki {\em et al.} from our 
formalism.  Finally,  in Section~\ref{sec_conclusions} we provide 
some conclusions and closing remarks.

\clearpage
\section{Steered Diffusion and Amplitude Amplification}

In this section we introduce the notion of {\em steering} 
operators in the context of Grover's algorithm.  These are 
operators that selectively bias either the diffusion process 
or the amplitude selection stage of the Grover algorithm,  
thereby providing for a vast array of amplitude amplification
strategies.  
Choosing the appropriate biasing scheme is of course the
critical issue in selecting a steering operator.  In a database 
search,  for example,  the correct biasing is determined solely 
by the details of the database itself.  In other instances,  when 
something is known {\em a priori} about the desired target 
state,  biasing can be used to increase the efficiency of the 
Grover search.  

\subsection{Notation and Context}
\label{sec_notation}

To establish some context and motivation,  let us consider a 
coherent $n$-qubit system.  Recall that the corresponding 
Hilbert space $\mathbb{H}_n$ has dimension $N=2^n$,
with computational basis states denoted by 
\begin{eqnarray}
  \vert x \rangle
  \equiv
  \vert x_{n-1} \,,\,  \cdots, x_1 \,, \,x_0 \rangle
  \equiv
  \vert x_{n-1} \rangle  \otimes \cdots \otimes \vert x_1 \rangle
  \otimes \vert x_0 \rangle 
  \,,
\label{eq_x_basis}
\end{eqnarray}
where $x_\ell \in \{0,1\}$ are binary observables labeled 
by the qubit indices $\ell \in \{0, \cdots, n-1\}$.   We shall 
denote the set of computational basis states for an 
$n$-qubit system by 
\begin{eqnarray}
  \Omega_n 
  \equiv
  \Big\{ \,  \vert x \rangle  \,\Big\vert\, x \in \{0,1\}^n \, \Big\}
  \ .
\label{eq_Omegan_def}
\end{eqnarray}
We have ordered the basis states such that the 
\hbox{$n$-tuple} \hbox{$x \equiv (x_{n-1},  \cdots,  x_1, x_0) 
\in \{0,1\}^n$} corresponds to the binary number $x_{n-1} 
\cdots x_1 \, x_0$,  where $x_0$ is the $2^0$-bit.  This is 
in keeping with OpenQASM syntax\,\cite{OpenQASM_ref}.   
For convenience,  we shall use a dual notation in which the 
binary number $x$ is represented by its corresponding 
base-10 number between $0$ and $N-1$.  In other words,
we shall index the computational basis state $\vert x \rangle$ 
by either a number 
$x \in  \{0,  1,  \cdots, N-1\}$,  or by the corresponding bit string 
$x \in\{0,1\}^n$.  Consequently,  we can express the set of 
basis elements in (\ref{eq_Omegan_def}) by either the index 
set $\Omega_n = \{0,  1,  \cdots,  N-1\}$,  or by the set of bit 
string $\Omega_n = \big\{x \,\vert\, x \in \{0,1\}^n \big\}$.  
In this dual notation,  we can list the computational basis 
elements  in the order given by their numerical base-10 
index,  
\begin{eqnarray}
  && \vert x_0 =0 \rangle \equiv \vert 0,  \cdots,  0, 0 \rangle
  \ ,
\\ \nonumber
  &&\vert x_1 =1 \rangle \equiv \vert 0,  \cdots,  0 ,1 \rangle
  \ ,
\\ \nonumber
  && \vert x_2 =2 \rangle \equiv \vert 0,  \cdots,  1, 0 \rangle
  \ ,
\\ \nonumber
  && \vert x_3=3 \rangle \equiv \vert 0,  \cdots,  1, 1 \rangle
  \ ,
\\ \nonumber
  && \hskip1.0cm \vdots \hskip2.5cm \vdots
\\ \nonumber
  &&\vert x_{\scriptscriptstyle N-1} =N-1 \rangle \equiv 
  \vert 1,  \cdots,  1, 1 \rangle
  \ .
\end{eqnarray}
Therefore,  if we wish to emphasize the order of a basis element,
we use the notation $\vert x_i \rangle$ with the index $i \in \{0,  
1,  \cdots,   N-1\}$.  This new index $i$ should not be confused 
with the qubit index $\ell \in \{0,  1,  \cdots,  n-1 \}$ of (\ref{eq_x_basis}),  
as they are used in different contexts.  

The traditional Grover's algorithm selects a single target
state $\vert \omega \rangle \in \Omega_n$,  amplifying this
component by the diffusion process.   Our goal is to generalize 
the algorithm to choose an arbitrary subset of basis states 
$\Omega \subseteq \Omega_n$.  We call the subset $\Omega$
 the {\em amplitude steering} set,  and the corresponding 
subspace $\mathbb{H}_\Omega \subseteq\mathbb{H}_n$ 
will be called the  steering subspace.  In 
general,  $\mathbb{H}_\Omega$ need not correspond to 
a multi-qubit system;  however,  it is of particular interest 
when it does,  and we shall investigate this case in some 
detail in a future section.  

Since the Hadamard vector $\vert h \rangle$ is so critical to 
the Grover algorithm,  we continue our analysis with a brief 
review of this state,  thereby providing continuity and establishing 
some notation.   The Hadamard state is defined by  
\begin{eqnarray}
  \vert h \rangle \equiv H^{\otimes n}\, \vert 0^n \rangle
  \ ,
\label{eq_had_two}
\end{eqnarray}
where $H$ is the single-qubit Hadamard gate and 
$\vert 0 \rangle^{\otimes n} \equiv \vert 0^n \rangle \equiv \vert 0, 
\cdots, 0, 0 \rangle$ is the $n$-qubit zero-state.  The 
Hadamard gate acts on the single-qubit computational 
basis states by
\begin{eqnarray}
  H \vert 0 \rangle
  &=&
  \frac{1}{\sqrt{2}} \Big[ \, \vert 0 \rangle  +  \vert 1 \rangle \Big]
  \equiv
  \vert + \rangle
\\[5pt]
  H \vert 1 \rangle
  &=&
  \frac{1}{\sqrt{2}} \Big[ \, \vert 0 \rangle  -  \vert 1 \rangle \Big]
  \equiv
  \vert - \rangle
  \ ,
\end{eqnarray}
where $\vert \pm \rangle$ are called the positive and negative 
Hadamard states,  respectively.  For an $n$-qubit system
we see that $\vert h \rangle  = \vert + \rangle^{\otimes n}$,  
and the Hadamard state can therefore be expressed as the 
uniform sum over all computational basis elements, 
\begin{eqnarray}
  \vert h \rangle
  =
  \frac{1}{\sqrt{2^n}} \sum_{x \in\{0,1\}^n} \vert x \rangle
  \ .
\label{eq_had_one}
\end{eqnarray}
Our dual qubit ordering convention allows us to change 
notation  to a more convenient form,  
\begin{eqnarray}
   \vert h \rangle
  =
  \frac{1}{\sqrt{N}} \sum_{x=0}^{N-1} \vert x \rangle
 \ ,
\label{eq_had_two}
\end{eqnarray}
and we shall move between the representations 
(\ref{eq_had_one}) and (\ref{eq_had_two}) at will.   

The Hadamard state $\vert h \rangle$ is used in the 
Grover algorithm in several ways.   First,  it serves as 
a trial or initial state upon which the amplitude and 
diffusion operators act.  Then the Hadamard state is 
passed to the phase oracle,  which marks the component 
of $\vert h \rangle$ in the direction of the target state 
$\vert \omega \rangle$ with a negative phase.  The 
resulting state is finally passed to the diffusion portion 
of the algorithm,  which amplifies the target component 
of the wave function.   The idea behind {\em steering} 
is that one can weight certain basis elements that are 
deemed more relevant,  replacing the uniformly 
weighted Hadamard state $\vert h \rangle$ in 
(\ref{eq_had_two}) with a biased trial state of the 
form 
\begin{eqnarray}
  \vert g \rangle 
  = 
   \sum_{x=0}^{N-1} g_x \, \vert x \rangle
  \equiv
  G \vert 0^n \rangle
  \ .
\label{eq_gen}
\end{eqnarray}
The operator $G$ is called the {\em diffusion steering 
operator}.  The weights $g_x$ are normalized complex 
numbers (although they are often taken to be real numbers),  
so that
\begin{eqnarray}
   \sum_{x=0}^{N-1} g_x \, g_x^* = 1
  \ .
\label{eq_gen_normal}
\end{eqnarray}
Since $\vert g \rangle$ is normalized to unity,  the steering 
operator must be unitary,  
{\em i.e.} $G^\dagger G = G G^\dagger =  \mathbb{1}$.  
Here,  $\mathbb{1}$ is the $2^n \times 2^n$ unit matrix 
for the $n$-qubit Hilbert space,  although if we wish 
to emphasize the number of qubits,   we shall write 
the unit operator as $\mathbb{1}_n$.  The diffusion 
operator corresponding to the state $\vert g \rangle$ 
is a simple Householder reflection,  
\begin{eqnarray}
   U_g 
  &\equiv&
 2 \vert g \rangle \langle g \vert - \mathbb{1}
\\[5pt]
  &=&
  G \Big[ \, 2 \vert 0^n \rangle \langle 0^n \vert - \mathbb{1} 
  \, \Big] G^\dagger
  \ .
\label{eq_Ug_def}
\end{eqnarray}
When we employ the Hadamard operator $H^{\otimes n}$ 
for the diffusion step,  we are using an unbiased linear 
superposition of computational basis states,  whereas the 
operator $G$ biases the diffusion in a well chosen manner.  
This will allow us to {\em steer} the diffusion process.  We 
can also bias the amplitude selection in a similar way
by expressing the target state as 
\begin{eqnarray}
  \vert \omega \rangle \equiv A_\omega \, \vert 0^n \rangle
  \ ,
\end{eqnarray}
where $A_\omega$ is a unitary operator called the {\em 
amplitude steering operator}.  The phase oracle that marks 
the state $\vert \omega \rangle$ is also a Householder 
reflection,  and takes the form 
\begin{eqnarray}
   U_\omega 
  &\equiv&
  \mathbb{1} - 2 \vert \omega \rangle \langle \omega \vert 
\\[5pt]
  &=&
  A_\omega \Big[ \, \mathbb{1} - 2 \vert 0^n \rangle \langle 
  0^n  \vert \, \Big] A_\omega^\dagger
  \ .
\label{eq_Uomega_def_one}
\end{eqnarray}
Employing well-chosen steering operators $G$ and 
$A_\omega$ can dramatically improve the efficiency and 
flexibility of the Grover search.  

Note that the diffusion and amplitude selection operators 
(\ref{eq_Ug_def}) and (\ref{eq_Uomega_def_one}) depend 
upon the operator $\mathbb{1} - 2 \vert 0^n \rangle \langle 
0^n \vert$,  which is a Householder reflection about the 
zero-state,  and it can be  implemented in the circuit model 
by a simple multi-control $Z$-gate.  The steering operators
$G$ and $A_\omega$,  
however,  require much more care.  They are generally (but 
not always) built from an exponential number of gates,  
thereby potentially degrading the quantum advantage 
of Grover's algorithm.  As previously noted, 
this  exponential gate problem  
has plagued the Grover search algorithm since its inception.  
However,  this problem has recently been addressed by 
Ref.~\cite{aae21},   which introduces the notion of {\em 
approximate amplitude encoding} (AAE) in which the 
steering operators are approximated by a parameterized 
shallow quantum circuit using only a polynomial number 
of gates.   A machine learning process is then employed,  
by which the steering operator can be approximated to 
any desired accuracy.  We will return to this point in a later 
section.

\subsection{Grover's Algorithm for a Single Basis Element}
\label{sec_grover_review}

\begin{figure}[t!]
\begin{minipage}[c]{0.4\linewidth}
\includegraphics[scale=0.23]{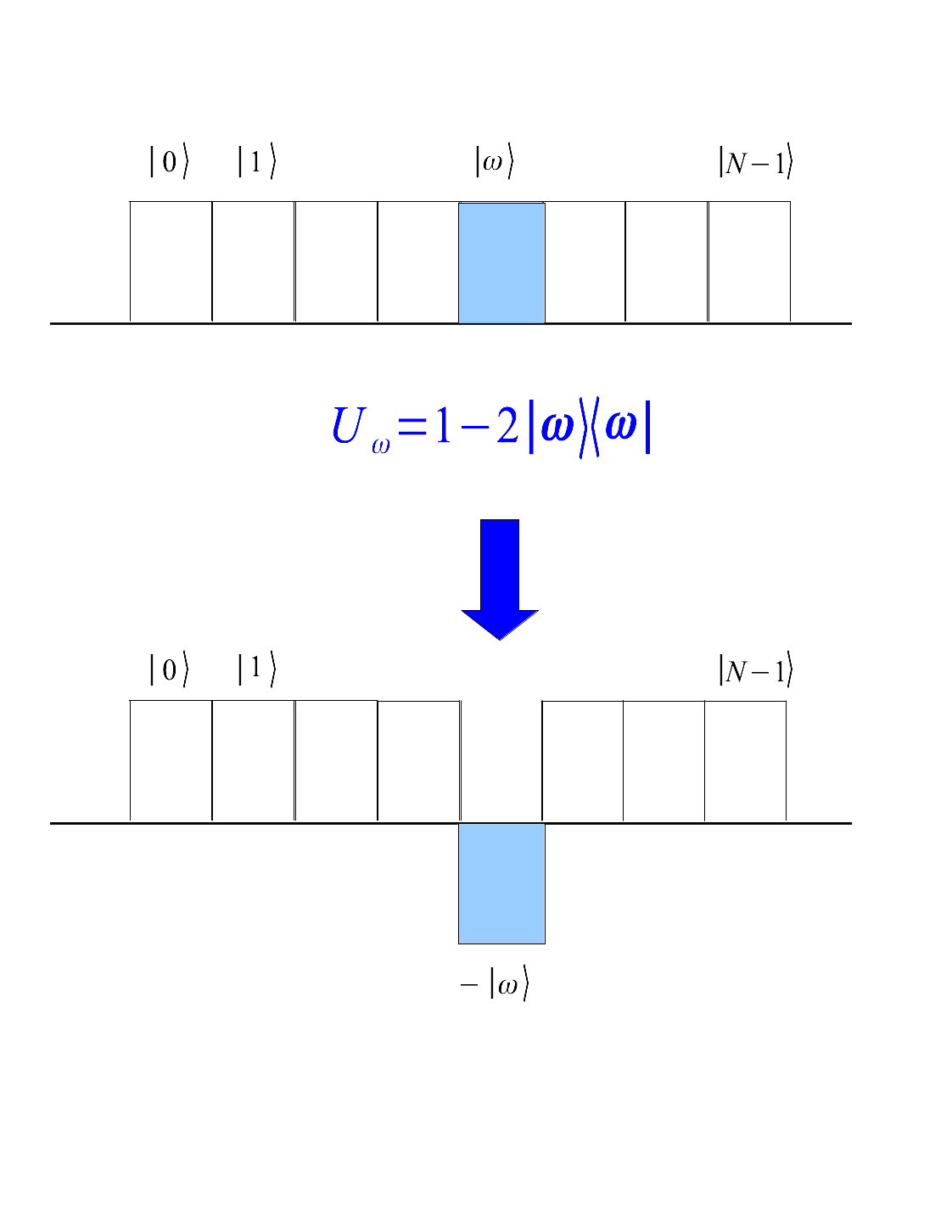}
\end{minipage}
\hfill
\begin{minipage}[c]{0.45\linewidth}
\includegraphics[scale=0.23]{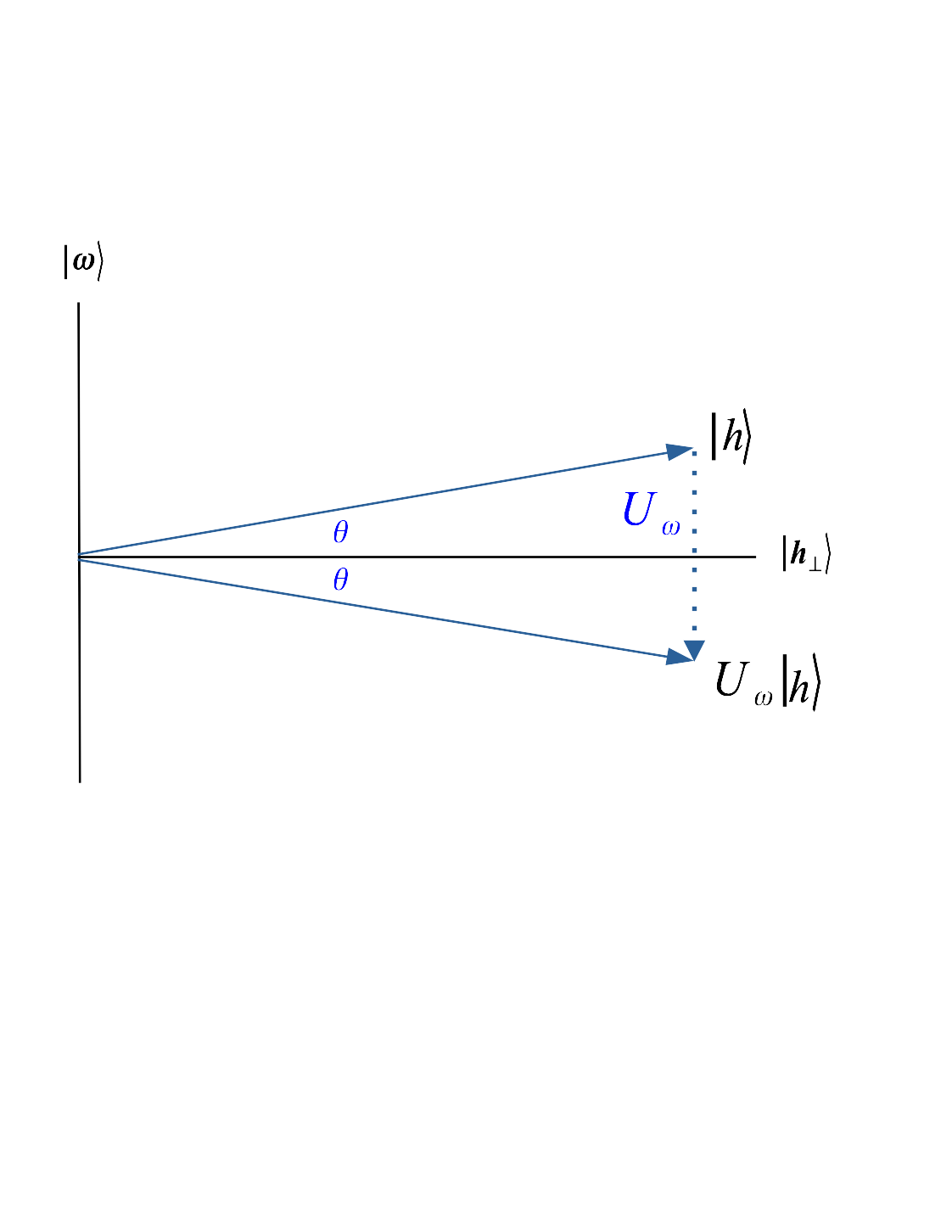}
\end{minipage}
\vskip-2.0cm
\caption{\footnoteskip
  Left panel: The phase oracle $U_\omega$ marks the target 
  state $\vert \omega \rangle$  with a negative phase, and therefore 
  the action of the oracle is given by the Householder reflection 
  $U_\omega = \mathbb{1} - 2 \vert \omega \rangle \langle \omega 
  \vert$. Right panel: The oracle is applied to the Hadamard state
  $\vert h \rangle$.  Note that $U_\omega$  inverts the component 
  of $\vert h \rangle$ in the direction of $\vert \omega \rangle$,  
  thereby reflecting $\vert h \rangle$ across the orthogonal 
  axis  $\vert h_\perp \rangle$.   If the angle between $\vert h 
  \rangle$ and $\vert h_\perp   \rangle$ is defined to be 
  $\theta$,  then the resulting state $U_\omega \vert h 
  \rangle$ lies below the orthogonal direction $\vert h_\perp  
  \rangle$ by the same angle $\theta$. 
}
\label{fig_marked_state}
\end{figure}

For continuity and completeness,   we now derive the original 
form of Grover's algorithm,  using well-motivated physical
arguments.  Our exposition will closely follow the presentation 
of Ref.~\cite{wiki_grover}.   Suppose we wish to  find a specific 
target state $\vert \omega \rangle$ from among the $N=2^n$ 
basis sates $\vert x \rangle$.  To proceed,  consider the {\em
unitary} operator $U_\omega$ defined on the basis states by
\begin{eqnarray}
   U_\omega \, | x \rangle 
  \equiv
  \left\{
  \begin{array}{ll}
   - | x \rangle & 
  {\rm for}~x=\omega 
  \\
  \phantom{-} | x \rangle & 
  {\rm for}~x \ne \omega \ ,
  \end{array}
  \right. 
\label{eq_U_omega_def}
 \end{eqnarray}
as illustrated in the left panel of  Fig.~\ref{fig_marked_state}.  
Although (\ref{eq_U_omega_def}) only defines $U_\omega$ 
on the basis states,  since the operator is also {\em linear},   its 
action is in fact defined on any state in the \hbox{$N$-dimensional} 
quantum Hilbert space  $\mathbb{H}_n$,  
\begin{eqnarray}
   U_\omega \, | \psi \rangle 
  = 
  U_\omega 
  \left(\, \sum_{x=0}^{N-1} \psi_x \, \vert x \rangle\right)
  =
  \sum_{x=0}^{N-1} \psi_x \,  U_\omega \,  
  \vert x \rangle
  \ .
\label{eq_Uomega_lin}
 \end{eqnarray}
The operator $ U_\omega$ is often called an {\em oracle} 
because it can be thought of as black-box that answers 
the question:  ``Which state should I chose?'' It is called 
a {\em phase oracle} because the answer is encoded by 
marking the {\em best} choice with a negative phase.  
In other words,   the purpose of the phase oracle $ U_\omega$ 
is to insert a negative phase on the target state  $\vert \omega 
\rangle$,  or on the component of the wave function in the target 
direction.  Note that we can express the phase oracle by a simple 
Householder reflection, 
\begin{eqnarray}
   U_\omega 
  = 
  \mathbb{1} - 2 \vert \omega \rangle \langle \omega \vert
  \ ,
\label{eq_Uomega_hh}
 \end{eqnarray}
since this satisfies  (\ref{eq_U_omega_def}) on the basis states 
$\vert x \rangle$.    From  (\ref{eq_Uomega_hh}) we see  that 
$U_\omega$ is indeed unitary,  $U_\omega^\dagger\, U_\omega 
= \mathbb{1}$,  and it is therefore a valid gate transformation 
operator.  Furthermore,  the operator is self-adjoint or Hermitian,  
$U_\omega^\dagger = U_\omega$,  and so that the unitary 
condition becomes $U_\omega^2 = \mathbb{1}$,  and we 
see that $U_\omega$ is also idempotent.   Also note that  
(\ref{eq_Uomega_hh}) implies that the action of $U_\omega$ 
on a general state is  given by 
\begin{eqnarray}
   U_\omega \vert \psi \rangle
  = 
  \vert \psi \rangle - \big(2\,\langle \omega \vert \psi \rangle\big)\, 
  \vert \omega \rangle 
  \  ,
 \end{eqnarray}
a more concise form than the one given in (\ref{eq_Uomega_lin}).  
We see explicitly that the component of the wave function in the
direction $\vert \omega \rangle$ receives a negative phase.   Also 
note that the vector $ U_\omega \vert \psi \rangle$ always lies in 
the \hbox{2-dimensional} subspace spanned by $\vert \psi \rangle$ 
and $\vert \omega \rangle$.  This fact will allow us to employ a 
2-dimensional Euclidean analogy to visualize the algorithm.
It is crucial to the Grover algorithm that the action of the oracle 
in the $N$-dimensional quantum  Hilbert space reduces to 
simple reflections in a 2-dimensional subspace of this 
(potentially quite large) Hilbert space. 

The Grover algorithm begins by letting the oracle $ U_\omega$ 
act on the Hadamard state $\vert h \rangle$,  thereby forming the
marked state $ U_\omega \vert h \rangle$.  Because $ U_\omega$ 
is a Householder 
reflection,  the vector $U_\omega \vert h \rangle$ will lie in the 
2-dimensional subspace spanned by  $\vert \omega \rangle$ 
and $\vert h \rangle$.  To visualize this reflection,  it is convenient 
to consider the state $\vert h_\perp \rangle$ orthogonal to 
$\vert \omega \rangle$ in this 2-dimensional subspace.   We 
can construct $\vert h_\perp \rangle$ by simply removing $\vert 
\omega \rangle$ from  $\vert h \rangle$,   and then  normalizing 
the resulting state, 
\begin{eqnarray}
  \vert h_\perp \rangle
  \equiv
  \frac{1}{\sqrt{N-1}} \sum_{x \ne \omega} \, \vert  x \rangle 
  \ .
 \end{eqnarray}
\begin{figure}[b!]
\includegraphics[scale=0.25]{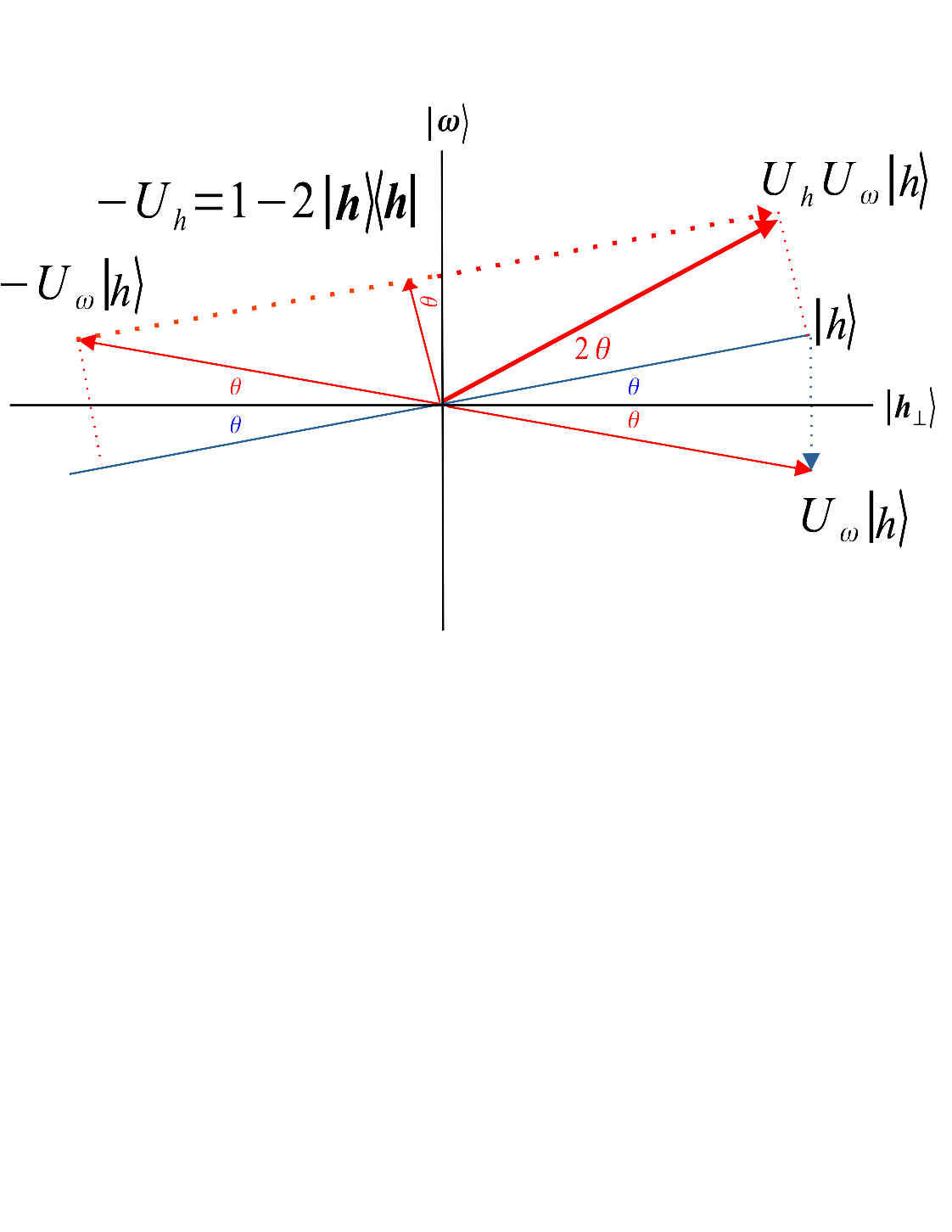} 
\vskip-6.0cm 
\caption{\footnoteskip  
The diffusion portion of the algorithm starts with the state 
$U_\omega \vert h \rangle$ in the fourth quadrant,  upon 
which we perform an inversion $-U_\omega \vert h \rangle$ 
into the second quadrant.  We then perform the Householder 
reflection $-U_h = \mathbb{1} - 2 \vert h \rangle \langle h \vert$ 
on $- U_\omega \vert h \rangle$,  placing $U_h \,  U_\omega 
\vert \omega \rangle$ into the first quadrant.   If the angle 
between $\vert h \rangle$ and $\vert h_\perp \rangle$ is 
$\theta$,  then the state $U_h \,  U_\omega \vert \omega 
\rangle$ will be inclined at an angle $2\theta$ above the 
original state $\vert h \rangle$,  and therefore $U_h \,  
U_\omega \vert h \rangle$ lies closer to the target vector 
$\vert \omega \rangle$ than does the starting position 
$\vert h \rangle$.  
}
\label{fig_flip_h}
\end{figure}
Note that $\langle h_\perp \vert h_\perp \rangle = 1$
and $\langle \omega \vert h_\perp \rangle = 0$, while
\begin{eqnarray}
  \langle h \vert h_\perp \rangle
  &=& 
 \sqrt{ \frac{N-1}{N}}
\\[5pt]
  \langle h \vert \omega \rangle
  &=& 
   \frac{1}{\sqrt{N}}
  \ .
 \end{eqnarray}
Let $\theta$ be the angle between $\vert h \rangle$ and 
$\vert h_\perp\rangle$ in the \hbox{2-dimensional} subspace,  
so that $\cos\theta  =  \langle h \vert h_\perp \rangle$ and 
$\sin\theta  =  \langle h \vert \omega\rangle$,  and we 
consequently find
\begin{eqnarray}
  \cos\theta
  =
 \sqrt{ \frac{N-1}{N}}
  ~~~\text{and}~~~
  \sin\theta
  = 
  \frac{1}{\sqrt{N}}
  \ .
  \label{eq_sin_theta}
 \end{eqnarray}
Since $ U_\omega$ is just a Householder reflection across the 
$\vert h_\perp \rangle$ hyperplane,   the state $U_\omega \vert h 
\rangle$ lies below the axis $\vert h_\perp \rangle$ at an angle of 
$\theta$,  as illustrated in the right panel of Fig.~\ref{fig_marked_state}.
For $\theta \ll 1$,  we have $\theta \approx 1/\sqrt{N}$.  


We are now ready for the next step of the algorithm,   the so called 
{\em diffusion} operation,  which is summarized in Fig.~\ref{fig_flip_h}.
In this step,  we start with the state $ U_\omega \vert h \rangle$ 
in the fourth quadrant  of the 2-dimensional subspace spanned 
by $\vert \omega \rangle$ and $\vert h_\perp \rangle$.  In other
words,  we start with $ U_\omega \vert h \rangle$ lying at an angle 
$\theta$ below the $\vert h_\perp \rangle$ axis.  The inverted 
state  $- U_\omega \vert h \rangle$ consequently lies in the second 
quadrant at an angle $\theta$ above the negative $\vert h_\perp
\rangle$ axis.   Let us define the diffusion operator
\begin{eqnarray}
  U_h \equiv 2 \vert h \rangle \langle h \vert - \mathbb{1} 
 \  ,
\label{eq_Uh_HHreflection}
 \end{eqnarray}
\begin{figure}[b!]
\includegraphics[scale=0.25]{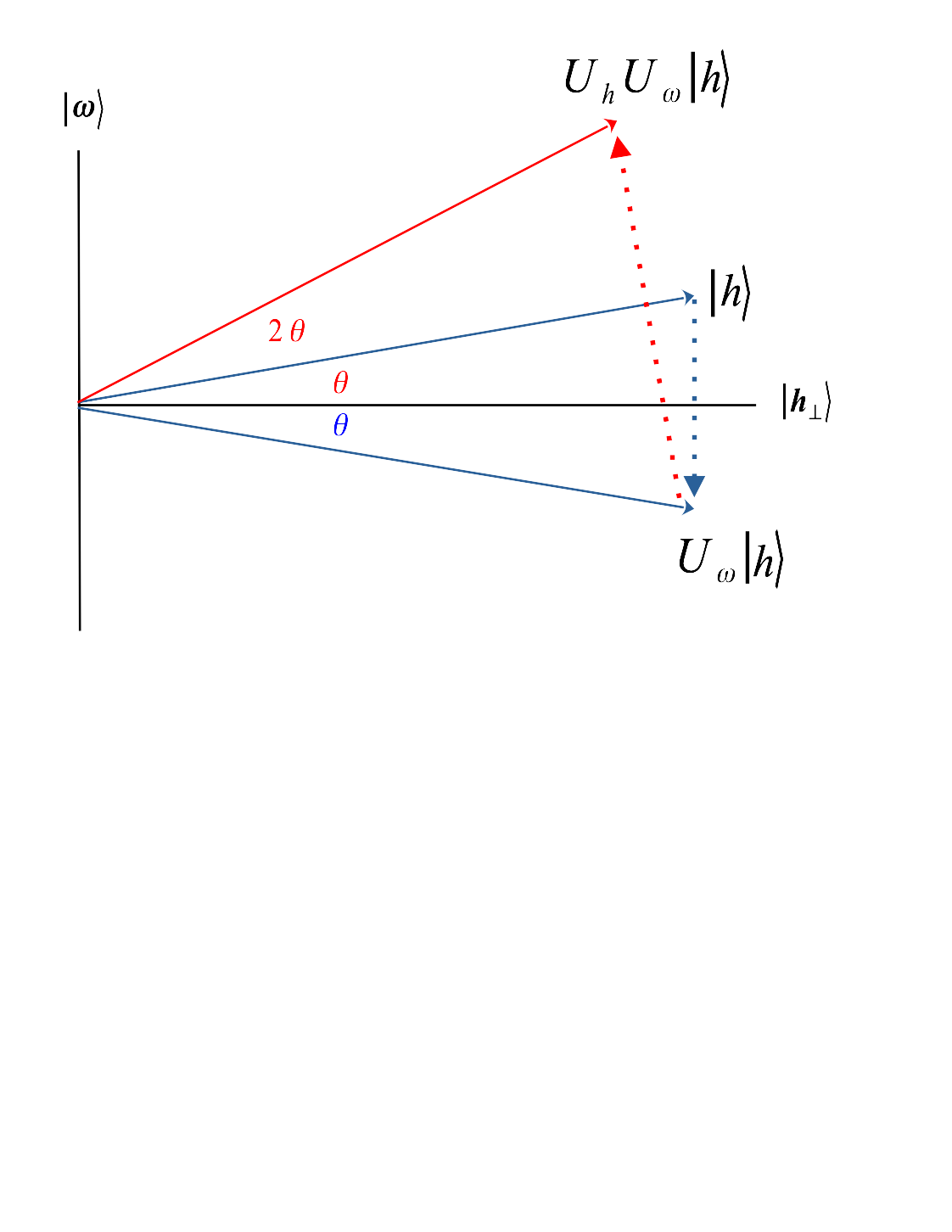} 
\vskip-6.0cm 
\caption{\footnoteskip The first Grover iteration.  We start 
with the Hadamard state $\vert h \rangle$ and apply a Householder 
reflection $ U_\omega = \mathbb{1} - 2 \vert \omega \rangle
\langle \omega \vert$ about the hyperplane orthogonal to 
$\vert \omega \rangle$.   This inverts the component of $\vert h 
\rangle$ that is parallel to $\vert \omega \rangle$ across the 
orthogonal direction $\vert h_\perp \rangle$,  as illustrated by 
the blue dashed arrow.  The state $U_\omega \vert h \rangle$ 
then undergoes a reflection $U_h =2\vert h \rangle \langle h 
\vert - \mathbb{1} $,  as illustrated by the red dashed line,  and this 
new state $U_h U_\omega \vert h \rangle$ lies closer to the 
target state $\vert \omega \rangle$  than the original vector 
$\vert h \rangle$.   After $r$ iterations,  the state  $(U_h U_\omega)^r 
\vert h \rangle$ is inclined at an angle $\theta_r = (2r+1)\theta$
above the horizontal axis $\vert h_\perp \rangle$,  and when 
$\theta_r= \pi/2$  we obtain the target state $\vert \omega
\rangle$ with unit probability.  For $\theta \ll 1$,  this corresponds 
to $r_* \approx (\pi/4)\sqrt{N}$ iterations.
}
\label{fig_grover_it}
\end{figure}

\vskip-0.5cm
\noindent
and we see that applying $-U_h =  \mathbb{1} - 2 \vert h \rangle 
\langle h \vert$ on $-U_\omega \vert h \rangle$ 
places $U_h U_\omega \vert h \rangle$ back into  the first quadrant.  
As the geometry of Fig.~\ref{fig_flip_h} reveals,   the resulting vector 
$U_h \,  U_\omega \vert h \rangle$ is inclined by angle $2\theta$ 
above the initial choice $\vert h \rangle$,  or an angle $3\theta$
above the horizontal axis $\vert h_\perp \rangle$.  Thus
$ U_h \,  U_\omega \vert h \rangle$ is closer to the target state 
$\vert \omega \rangle$  than the initial vector $\vert h \rangle$. 
Upon repeating these two operations,  $U_\omega$ followed by
$U_h$,  we obtain vectors that lie closer and closer to $\vert
\omega \rangle$.  The first such Grover iteration is summarized 
in Fig.~\ref{fig_grover_it}.  

We must be careful, however,   concerning the number of
iterations that we employ,  as it is possible to overshoot the 
target vector $\vert \omega \rangle$.  So how many iterations 
must we perform before we reach the target $\vert \omega
 \rangle$?  After $r$ iterations, the angle between $\vert h_\perp 
\rangle$ and the iterated state $\big(U_h \,  U_\omega\big)^r 
\vert h \rangle$ is $\theta_r = (2r+1) \theta$.  When $\theta_r 
= \pi/2$,   then we have performed enough iterations to extract 
the target state $\vert \omega \rangle$ with unit probability.  
For  $\theta \ll 1$, we must perform $r=r_*$ iterations until 
$2r_* \theta \approx \pi/2$,  which implies that the number 
of requisite iterations is $r_* \approx \pi/4\theta \approx (\pi/4)
\sqrt{N}$.  The fact that the proportionality constant is $\pi/4 
\approx 0.75 < 1$  is quite fortunate,  since a factor like $2\pi 
\approx 6$ would have required proportionally more iterations.

\begin{figure}[b!]
\includegraphics[scale=0.25]{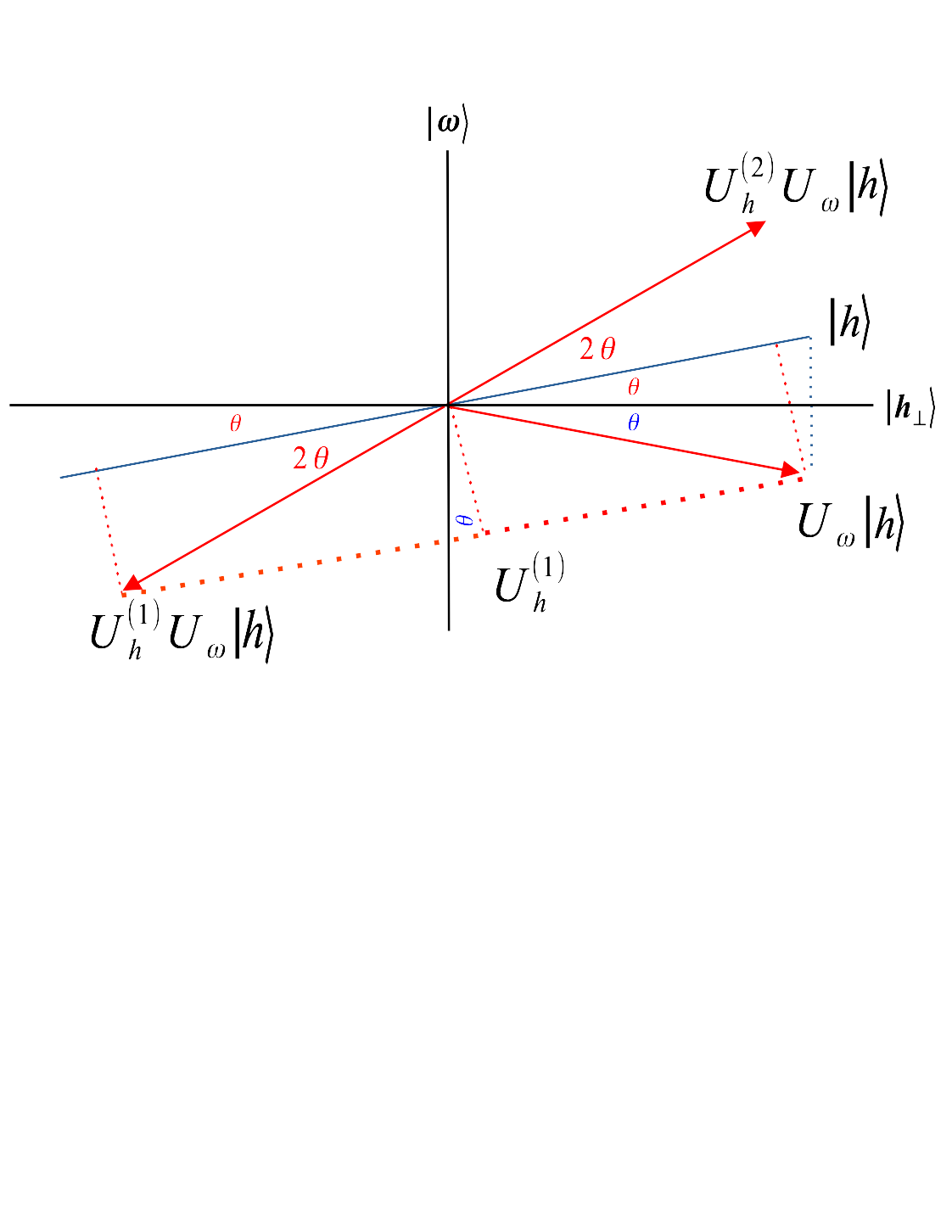} 
\vskip-6.0cm 
\caption{\footnoteskip  
Since the operators
$U_h^{(1)} = \mathbb{1} - 2 \vert h \rangle \langle h \vert$ 
and $U_h^{(2)} =2 \vert h \rangle \langle h \vert - \mathbb{1}$
differ by a sign,  the states $U_h^{(1)} U_\Omega \vert h 
\rangle$ and $U_h^{(2)} U_\Omega \vert h \rangle$ are 
physically indistinguishable.  
}
\label{fig_flip_minus_h}
\end{figure}
There is a subtlety that we should mention regarding an overall 
negative phase factor in Fig.~\ref{fig_flip_h}. Quantum mechanics 
cannot distinguish between the action of the two operators 
$U_h^{(1)} = \mathbb{1} - 2 \vert h \rangle \langle h \vert$ and 
$U_h^{(2)} =2 \vert h \rangle \langle h \vert - \mathbb{1}$ since 
they differ by a sign (a total phase of $e^{i \pi}$).  Recall that 
we first acted on the state $\vert h \rangle$ with the operator 
$U_\omega$,  so that $U_\omega \vert h \rangle$ lies in the 
fourth quadrant.  We then  inverted the vector to place 
$-U_\omega \vert h \rangle$ into the second quadrant.   
However,   the two states $ U_\omega \vert h \rangle$ and 
$- U_\omega \vert h \rangle$ are indistinguishable,  so we 
might as well act on $ U_\omega \vert h \rangle$  with the 
operator $U_h^{(1)} = \mathbb{1} - 2 \vert h \rangle \langle h 
\vert$,  giving a resultant vector $ U_h^{(1)} \,  U_\omega 
\vert h \rangle$ in the third quadrant,  as illustrated in 
Fig.~\ref{fig_flip_minus_h}.  The negative of this vector,   
$- U_h^{(1)} \,  U_\omega \vert h \rangle$,  is just the final 
vector $ U_h^{(2)}\,  U_\omega \vert h \rangle$ in the first 
quadrant of Fig.~\ref{fig_flip_h}.
But since quantum mechanics is insensitive to an overall 
phase,  both $ U_h^{(1)}$ and $ U_h^{(2)}$ are represented 
by the same quantum circuit.   We therefore use $U_h \equiv 
U_h^{(2)} = 2\vert h \rangle \langle h \vert - \mathbb{1}$ in 
the text of this manuscript,  while we employ $- U_h
= U_h^{(1)}  = \mathbb{1} - 2\vert h \rangle \langle h \vert$ 
in quantum circuits.  This is because the quantum
circuits for $U_h^{(1)}$ and $U_h^{(2)}$ differ by
a minus sign,  and the circuit implementation of
$-U_h=\mathbb{1} - 2\vert h \rangle \langle h \vert$
and $U_\omega = \mathbb{1} - 2\vert \omega \rangle 
\langle \omega \vert$ will be quite similar.

\subsection{Grover's Algorithm with Steering}
\label{sec_grover_steering}

\begin{figure}[h!]
\includegraphics[scale=0.25]{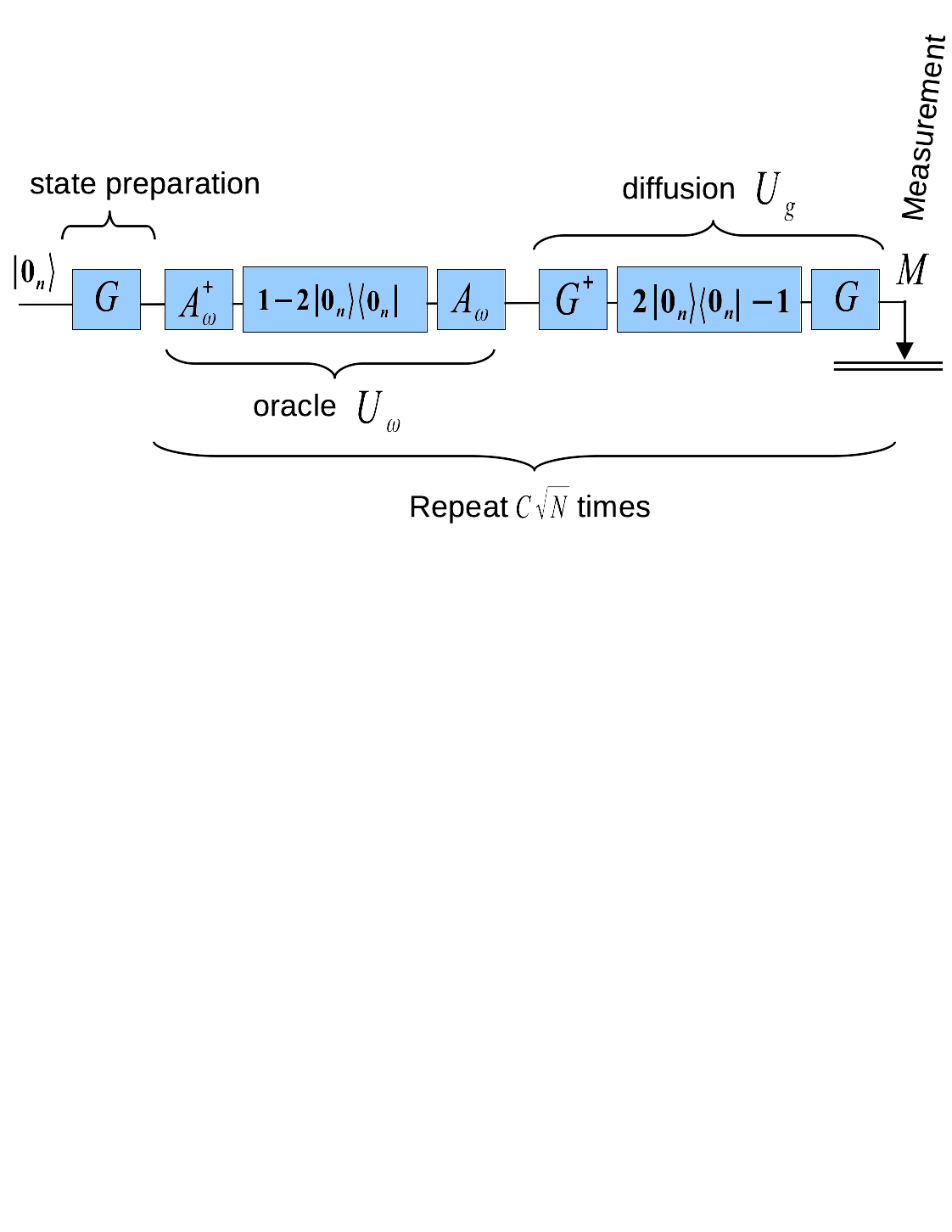} 
\vskip-7.0cm 
\caption{\footnoteskip  
 Steered Grover's algorithm
}
\label{fig_grover_algorithm}
\end{figure}

In the previous section,  we can replace the target basis 
state $\vert\omega \rangle \in \Omega_n$ by a general 
target state
\begin{eqnarray}
  \vert \omega \rangle 
  =
  \sum_{x=0}^{N-1} \omega_x \vert x \rangle 
  \in \mathbb{H}_n 
  \ ,
 \end{eqnarray}
and the argument will remain the same.  We can write 
$\vert\omega \rangle =A_\omega \, \vert 0^n \rangle$,  
where $A_\omega$ is the unitary amplitude steering 
operator,  and the oracle takes the form
\begin{eqnarray}
  U_\omega 
  &=& \mathbb{1} - 2 \vert \omega \rangle \langle \omega \vert
\\[5pt]
  &=& 
  A_\omega \Big[ \mathbb{1} - 2 \vert 0^n \rangle \langle 0^n 
  \vert\, \Big] 
  A_\omega^\dagger
  \ .
 \end{eqnarray}
The operator $\mathbb{1} - 2 \vert 0^n \rangle \langle 0^n \vert$ 
marks the zero-state,  and in the quantum circuit model it can be
implemented by a multi-control $Z$-gate.
The argument also goes through unscathed if we replace the 
Hadamard state $\vert h \rangle$ with the steered state $\vert 
g \rangle = G \vert 0^n \rangle$ of (\ref{eq_gen}).  That is to say,  
let us start with the trail state $\vert g \rangle$,  apply the oracle 
$U_\omega$,  and then  steer the diffusion with the operator
\begin{eqnarray}
  U_g 
  &=&
  2 \vert g \rangle \langle g \vert - \mathbb{1} 
\\[5pt]
  &=& 
  G \Big[ 2 \vert 0^n \rangle \langle 0^n \vert\,  - \mathbb{1} 
  \Big] G^\dagger
  \ .
\label{eq_Ug_HHreflection_first}
 \end{eqnarray}
The form of the steered diffusion operator 
(\ref{eq_Ug_HHreflection_first}) is a special case of 
Hadamard diffusion, 
\begin{eqnarray}
  U_h
  =
  2 \vert h \rangle \langle h \vert - \mathbb{1} 
  =
  H^{\otimes n } \Big[
  2\vert 0^n \rangle \langle 0^n - \mathbb{1} \vert
  \Big] H^{\otimes n }
  \ . 
 \end{eqnarray}
Note that the operator $\mathbb{1} - 2\vert 0^n \rangle \langle 0^n 
\vert$ is an oracle that marks the state $\vert 0^n \rangle$ with 
a negative phase.   The operators $U_\omega$,  $ U_h$,  and 
$U_g$ can be constructed from a combination of multi-controlled 
$Z$-gates and single-qubit $X$ and $H$ gates.
Let $\theta$ be the angle of inclination of $\vert g\rangle$,
and for $\theta \ll 1$,  then the number of Grover iterations 
required to pull out the target state is 
\begin{eqnarray}
  r_*  \approx \frac{\pi}{4\theta} \approx C  \sqrt{N}
  \ ,
\end{eqnarray}
where the prefactor $C$ depends upon the choice of the 
steering operator  $G$.   When we take $G = H^{\otimes n}$,  
then the prefactor reduces to its previous value $C_h \equiv 
\pi/4$.  For a well-chosen $G$,  the prefactor $C$ can be much 
smaller than $C_h$,  and fewer iterations will be required.   We 
therefore arrive at the following algorithm,  which is illustrated 
in Fig.~\ref{fig_grover_algorithm}.

\clearpage
\vskip0.5cm
\centerline{\bf The Algorithm}
\vskip0.2cm
\begin{enumerate}
  \baselineskip 10pt plus 1pt minus 1pt
  \setlength{\itemsep}{3pt} 
  \setlength{\parskip}{1pt} %
  \setlength{\parsep}{0pt}  %
\item[1.] Initialize the system to the state
\begin{eqnarray}
  \vert g \rangle 
  =
  G \vert 0 \rangle^{\otimes n}
  =
  \sum_{x=0}^{N-1} g_x \vert x \rangle 
  \ ,
\label{eq_g_state_def}
 \end{eqnarray}
and construct the target state
\begin{eqnarray}
  \vert \omega \rangle 
  =
  A_\omega \vert 0 \rangle^{\otimes n}
  =
  \sum_{x=0}^{N-1} \omega_x \vert x \rangle 
  \ .
\label{eq_a_state_def}
 \end{eqnarray}
\item[2.] Perform a {\em Grover iteration} on the state vector:
\begin{enumerate}
  \baselineskip 10pt plus 1pt minus 1pt
  \setlength{\itemsep}{3pt} 
  \setlength{\parskip}{1pt} %
  \setlength{\parsep}{0pt}  %
  \item[(a)] Apply the phase oracle $ U_\omega = \mathbb{1} 
  - 2 \vert \omega \rangle  \langle \omega \vert $,  
   where $\vert \omega \rangle = A \, \vert 0 \rangle^{\otimes n}$. 
  \item[(b)] Apply the diffusion operator $U_g = 2 \vert g \rangle 
  \langle g \vert - \mathbb{1}$,  where $\vert g \rangle = G \, 
  \vert 0 \rangle^{\otimes n}$. 
\end{enumerate}
\item[3.] Repeat this of order $r_* \approx \pi/4\theta \approx
  C \sqrt{N}$ times. 
\item[4.] Measure the resulting state  in the computational basis 
after $r_*$ iterations.  The output will be $\vert\omega\rangle$ 
with probability approaching unity.  The numerical value of $C$ 
depends upon the steering operator $G$. 
\end{enumerate}
\clearpage
\subsection{Grover's Algorithm for an Arbitrary Target Set}
\label{sec_grover_general}

In the last section,  we devised a phase oracle $U_\omega$ 
that marks any given computational basis state $\vert\omega 
\rangle$ from among the $N=2^n$ possible basis states of 
an $n$-qubit system $\mathbb{H}_n$.   That is to say,  the 
phase oracle $U_\omega$ marks a {\em single} target vector 
$\omega \in \Omega_n$,   where the set of all  computational 
basis states is represented by \hbox{$\Omega_n = \big\{ x \,
\big\vert\, x \in \{0,1\}^n  \big\}$}.   In this section 
we construct a phase oracle $U_\Omega$ that marks an 
arbitrary subset of target vectors  \hbox{$\Omega \subseteq 
\Omega_n$}.  We shall call $\Omega$ the {\em steering set},  
and assume that it contain $M \le N$ elements.   The 
Hilbert space associated with $\Omega$ will be denoted 
$\mathbb{H}_\Omega$,  and called the steering subspace,
while the orthogonal Hilbert space will be denoted $\mathbb{H
}_\perp$.  The space $\mathbb{H}_\Omega$ need not be a 
multi-qubit subspace of $\mathbb{H}_n$,  although this is a 
very interesting case that we will shortly consider.  

Recall that we are using a dual notation in which a computational 
basis vector can be indexed by either a binary number (or bit string) 
$x \in \{0,1\}^n$,  or by the corresponding \hbox{base-10} 
number $x \in \{0,  1,  \cdots,  N-1\}$.  Thus,  if we wish 
to enumerate the basis elements,  we employ the notation 
\hbox{$\Omega_n  = \big\{ x_0,  x_1,  \cdots,  x_{N-1} \big\}$}
in which $x_i$ is the \hbox{base-10} representation of the 
basis element $i \in \{0,  1,  \cdots,  N-1\}$.   We can therefore 
express the set of basis elements by the notation 
$\Omega_n =\big\{0,  1,  \cdots,  N-1\big\}$,  or even by the 
collection of ket vectors $\Omega_n=\big\{ \vert x_0 \rangle,  
\vert x_1 \rangle,  \cdots,  \vert x_{N-1}  \rangle \big\}$,  depending 
upon on the context.   We will not necessarily choose the basis 
elements of $\Omega$ in any specific order,  so we express 
the steering set by $\Omega = \big\{ x_{i_0} ,  x_{i_1} ,  \cdots,  
x_{i_{M-1}} \big\}$.  Adopting the notation $\omega$ to 
reference a target element,  we can also write $\Omega = 
\big\{ \omega_{0} ,  \omega_{1},  \cdots,  \omega_{{M-1}} \big\}$,   
or more precisely,  $\Omega = \big\{ \vert \omega_{0}\rangle,  
\vert\omega_{1} \rangle ,  \cdots,  \vert\omega_{{M-1}}  \rangle  
\big\}$.  Note that there are  $2^N = 2^{2^n}$ subsets of
$\Omega_n$ (including the empty set),  each corresponding
to a phase oracle $U_\Omega$. 

\subsubsection{Marking a Subset of the Computational Basis}

\begin{figure}[b!]
\includegraphics[scale=0.25]{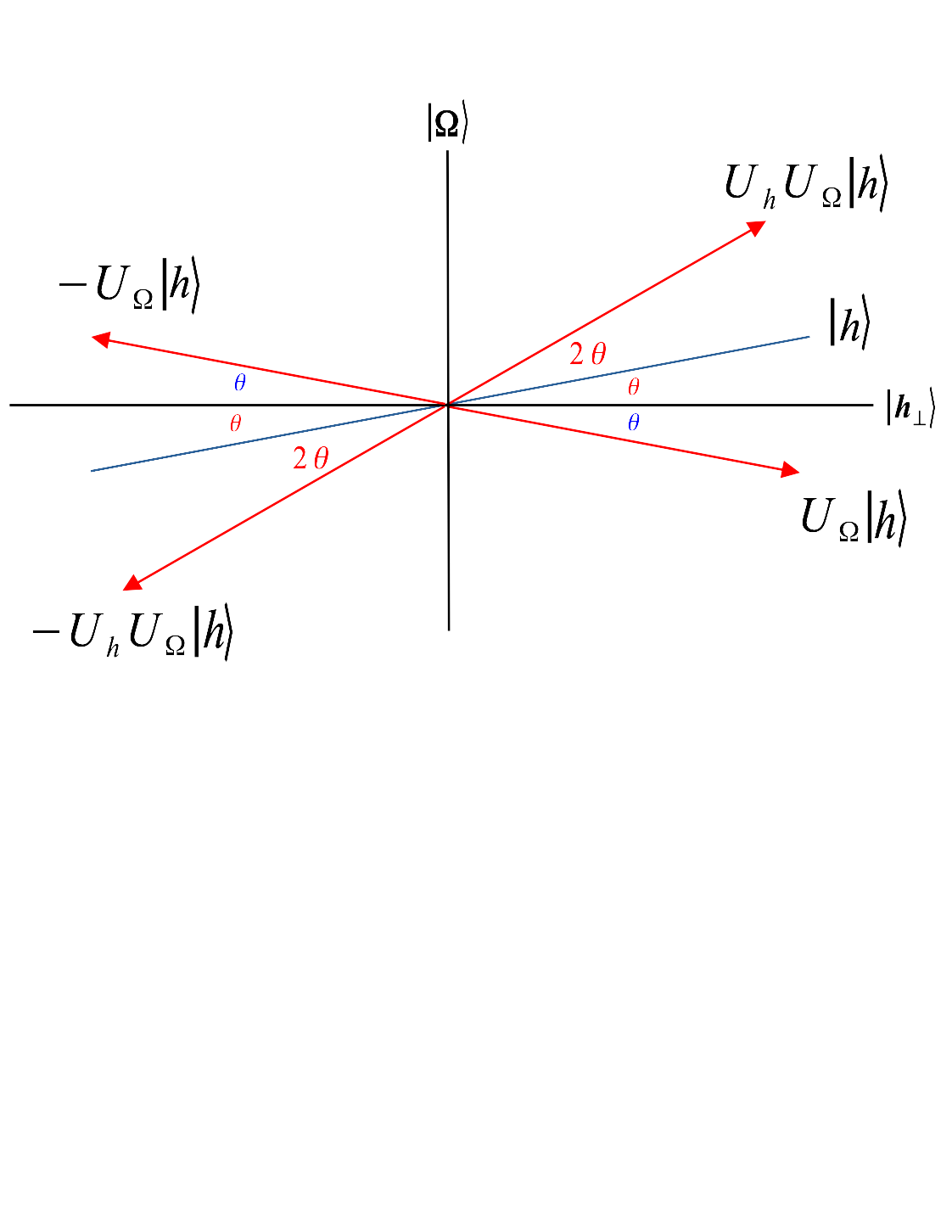} 
\vskip-6.0cm 
\caption{\footnoteskip  
  The action of $U_\Omega=\mathbb{P}_\perp - \mathbb{P}_\Omega$ 
  and $U_h = 2 \vert h \rangle \langle h \vert - {\,\sf 1} $  remain in the 
  \hbox{2-dimensional} subspace spanned by $\vert \Omega \rangle$ 
  and $\vert h_\perp\rangle$.  The state $\vert h \rangle$ lies within 
  this subspace,  and is inclined at an angle $\theta$  above 
  $\vert h_\perp\rangle$.  The  state $U_\Omega \vert h \rangle$ lies 
  below the $h_\perp$ axis at an angle $\theta$,   while the iterated 
  state $U_h U_\Omega \vert h \rangle$ is inclined at an angle $2\theta$ 
  above $\vert h\rangle$,  or $3\theta$ above $\vert h_\perp\rangle$.
}
\label{fig_flip_h_Omega}
\end{figure}

The phase oracle $U_\Omega$ marks all basis states in the 
steering set $\Omega\subseteq \Omega_n$ with a negative
phase,  and its action on a general $x \in \Omega_n$ is therefore
given by 
\begin{eqnarray}
   U_\Omega \, | x \rangle 
  &=&
  \left\{
  \begin{array}{ll}
   - | x \rangle &  {\rm for}~ x \in \Omega 
\\
  \phantom{-} | x \rangle &  {\rm for}~ x \notin \Omega \ .
  \end{array}
  \right. 
\label{eq_UOmega_def_A}
 \end{eqnarray}
Consequently,  we can express the oracle by
\begin{eqnarray}
   U_\Omega 
  &=&
   \mathbb{1} - 2  \sum_{x\in \Omega} 
  \vert x \rangle \langle x \vert 
  \ ,
\label{eq_UOmega_HH_fx}
 \end{eqnarray}
since (\ref{eq_UOmega_HH_fx}) satisfies (\ref{eq_UOmega_def_A}) 
for every basis element  $\vert x \rangle$.  We now decompose the 
unit operator in terms of basis state projection operators,  
\begin{eqnarray}
   \mathbb{1} 
  =
  \sum_{x=0}^{N-1}  \vert x \rangle \langle x \vert 
  =
  \sum_{x \notin \Omega}  \vert x \rangle \langle x \vert 
  +
  \sum_{x \in \Omega}  \vert x \rangle \langle x \vert 
  =
  \mathbb{P}_\perp + \mathbb{P}_\Omega
  \ ,
 \end{eqnarray}
where the projection operators onto $\mathbb{H}_\Omega$
and the orthogonal subspace $\mathbb{H}_\perp$ are defined 
by 
\begin{eqnarray}
  \mathbb{P}_\Omega
  &=&
  \sum_{x \in \Omega}  \vert x \rangle \langle x \vert 
\\[5pt]
 \mathbb{P}_\perp
  &=&
  \sum_{x\notin \Omega}  \vert x \rangle \langle x \vert 
  \ .
 \end{eqnarray}
The phase oracle (\ref{eq_UOmega_HH_fx}) now
takes a particularly simple form,  
\begin{eqnarray}
   U_\Omega 
  &=&
  \sum_{x\notin \Omega}   \vert x \rangle \langle x \vert 
  - 
  \sum_{x\in \Omega}   \vert x \rangle \langle x \vert 
  =
  \mathbb{P}_\perp - \mathbb{P}_\Omega
  \ .
\label{eq_UOmega_Proj}
 \end{eqnarray}
The action of the phase oracle on a general state is therefore 
\begin{eqnarray}
   U_\Omega \, | \psi \rangle 
  = 
   U_\Omega 
  \left( \,  \sum_{x=0}^{N-1} \psi_x \, \vert x \rangle\right)
  =
  \sum_{x \notin \Omega}  \psi_x \, \vert x \rangle
  -
  \sum_{x \in \Omega}  \psi_x \, \vert x \rangle
  \ ,
\end{eqnarray}
or more specifically,  its action on the Hadamard state is
\begin{eqnarray}
   U_\Omega \, | h \rangle 
  =
  \frac{1}{\sqrt{N}}\sum_{x \notin \Omega}  \, \vert x \rangle
  -
  \frac{1}{\sqrt{N}}\sum_{x \in \Omega}   \, \vert x \rangle
  \ .
\label{eq_UOmega_on_h}
\end{eqnarray}
This suggests that we consider the \hbox{2-dimensional} 
subspace spanned by the orthonormal states
\begin{eqnarray}
  \vert \Omega \rangle
  &\equiv&
  \frac{1}{\sqrt{M}} \sum_{x \in \Omega} \vert x \rangle
\label{eq_Omega}
\\[5pt]
  \vert h_\perp \rangle
  &\equiv&
  \frac{1}{\sqrt{N-M}} \sum_{x \notin \Omega} \vert x \rangle
  \ .
\label{eq_hPerp}
 \end{eqnarray}
One can easily check that the states $\vert \Omega \rangle$ 
and $\vert h_\perp \rangle$ are orthogonal and normalized
to unity,   {\em i.e.}  $\langle \Omega \vert \Omega \rangle
=\langle h_\perp \vert h_\perp \rangle = 1$,  and 
$\langle \Omega \vert h_\perp \rangle = 0$.  Also note 
that $\vert \Omega \rangle$ and $\vert h_\perp \rangle$ 
have the following overlaps with the Hadamard state 
$\vert h \rangle$,  
\begin{eqnarray}
  \langle \Omega \vert h \rangle
  &=&
  \sqrt{\frac{M}{N}} 
  ~~~\text{and}~~~
  \langle h_\perp \vert h \rangle
  =
  \sqrt{\frac{N-M}{N}}
  \ .
\label{eq_hhper_inner}
 \end{eqnarray}
We shall call 
the \hbox{2-dimensional} space spanned by  $\vert \Omega 
\rangle$ and $\vert h_\perp \rangle$ the $\Omega$-$h$ 
subspace.   The  Hadamard state $\vert h \rangle$ lies within 
this subspace,   and using relations (\ref{eq_hhper_inner}),  it 
can be decomposed as
\begin{eqnarray}
  \vert h \rangle
  =
  \sqrt{\frac{N-M}{N}} \, \vert h_\perp \rangle
  +
  \sqrt{\frac{M}{N}} \, \vert \Omega \rangle
  \ .
\label{eq_h_expand}
 \end{eqnarray}
Furthermore,  the operation of the phase oracle $U_\Omega$ 
on the Hadamard state remains within this subspace,  as applying 
(\ref{eq_UOmega_Proj}) to (\ref{eq_h_expand}) gives
\begin{eqnarray}
   U_\Omega \, | h \rangle 
  =
  \sqrt{\frac{N- M}{N}} \,\vert h_\perp \rangle
  -
  \sqrt{\frac{M}{N}}\, \vert \Omega \rangle
  \ .
\label{eq_UOmega_on_h_again}
\end{eqnarray}
This expression can also be obtained directly from 
(\ref{eq_UOmega_on_h}).  Obviously $-U_\Omega
\, \vert h \rangle$ also lies within the $\Omega$-$h$ 
subspace,  as does the action of the Householder reflection 
$U_h = 2 \vert h \rangle \langle h \vert -  \mathbb{1}$
on $U_\Omega \vert h \rangle$,  
\begin{eqnarray}
  U_h \,  U_\Omega \, \vert h \rangle
  &=&
  \Bigg[2 \vert h \rangle \langle h \vert  - \mathbb{1} 
  \Bigg]
  \Bigg[
  \sqrt{\frac{N-M}{N}} \, \vert h_\perp \rangle
  -
  \sqrt{\frac{M}{N}} \, \vert \Omega \rangle
  \Bigg]
\\[5pt]
   &=&
   \sqrt{\frac{N-M}{N}}\left[1 - \frac{4M}{N} \right] \vert h_\perp \rangle
  + 
  \sqrt{\frac{M}{N}}\left[ 3 - \frac{4M}{N} \right] \vert \Omega \rangle
  \ .
\label{eq_Uh_UOmega_gen}
 \end{eqnarray}
Further iterations $(\pm U_\Omega \, U_h)^r \vert h \rangle$ 
for $r \ge 1$ also lie within the $\Omega$-$h$ subspace.  
These results are illustrated in Fig.~\ref{fig_flip_h_Omega},
and provide the basis for a generalized Grover algorithm. 

\begin{figure}[h!]
\includegraphics[scale=0.60]{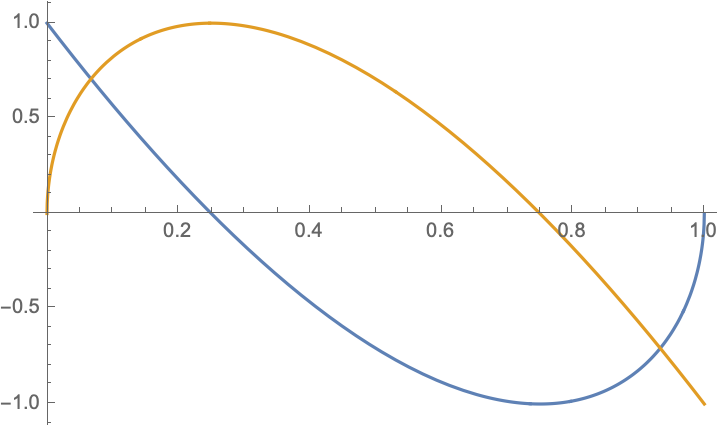} 
\caption{\footnoteskip  
  The coefficients of $U_h U_\Omega \vert h \rangle$
  from (\ref{eq_Uh_UOmega_gen}) as a function of $M/N$.  
  The blue line is the coefficient of $\vert h_\perp\rangle$ 
  and the yellow line is the coefficient of $\vert \Omega \rangle$. 
  This partitions the ratio $M/N$ into three regions: (i) $0 \le M/N
  \le 1/4$,  (ii) $1/4 < M/N  < 3/4$,  and $3/4 \le M/N  \le 1$.
  Regions (i) and (iii) permit the usual Grover algorithm,  while
  region (ii) requires an extra reflection. 
}
\label{fig_coeffs}
\end{figure}

To construct the algorithm,   let $\theta$ be the angle between 
$\vert h \rangle$ and $\vert h_\perp \rangle$,  so that
\begin{eqnarray}
  \cos\theta
  =
  \langle h_\perp \vert h \rangle
  = 
 \sqrt{ \frac{N-M}{N}}
  ~~~ \text{and}~~~
  \sin\theta
  =
  \langle \Omega \vert h \rangle
  = 
  \sqrt{\frac{M}{N}}
  \ ,
  \label{eq_sin_theta_gen}
 \end{eqnarray}
and we then expand the Hadamard state as
\begin{eqnarray}
  \vert h \rangle 
  = 
  \cos\theta \, \vert h_\perp\rangle
  + 
  \sin\theta \, \vert \Omega\rangle 
  \ .
\end{eqnarray}
We see that $\vert h \rangle$ lies in the 1-st quadrant of the 
\hbox{2-dimensional} $\Omega$-$h$ space at an 
angle $\theta$ above the $\vert h_\perp\rangle$ axis,  while
$$
  U_\Omega \vert h \rangle
  = 
  \cos\theta\, \vert h_\perp\rangle
  -
  \sin\theta \, \vert \Omega\rangle \ ,
$$
lies in the 4-th quadrant at an angle $\theta$ below the 
$\vert h_\perp\rangle $ axis.  Note,   however,  that 
\begin{eqnarray}
  U_h U_\Omega \vert h \rangle
  &=& 
  \cos\theta \Big[ 1 - 4 \sin^2\theta \Big] 
  \vert h_\perp\rangle
  -
  \sin\theta \Big[ 3 - 4 \sin^2\theta \Big]  
  \vert \Omega\rangle 
\\[5pt]
  &=& 
  \cos3\theta \,   \vert h_\perp\rangle 
  +
  \sin3\theta \,   \vert \Omega\rangle 
\end{eqnarray}
lies back in the 1-st quadrant only when the coefficients of 
$\vert h_\perp \rangle$ and $\vert \Omega \rangle$ are
{\em both} positive,  in which case it is inclined by $2\theta$
above $\vert h \rangle$,  or $3\theta$ above the horizontal
axis $\vert h_\perp \rangle$.  From (\ref{eq_Uh_UOmega_gen}) 
this corresponds to the region $0 \le M/N \le 1/4$.   If the 
coefficients are both negative,  so that $3/4 \le M/N \le 
1$,   then the Grover iteration places the state $U_h U_\Omega 
\vert h \rangle$ in the 3-rd quadrant,  which is equivalent 
to being in the 1-st quadrant (by a negative inversion).  
For $1/4 < M/N < 3/4$ the coefficients have different 
signs.  In fact,  the $\vert h_\perp \rangle$ coefficient
is always negative and the  $\vert \Omega \rangle$ 
coefficient is positive,  placing the state in the 2-nd 
quadrant.  In this case,  another Householder reflection 
of the form
\begin{eqnarray}
  U_\parallel 
  = 
  2 \vert \Omega \rangle \langle \Omega \vert 
  -
  \mathbb{1} 
\label{eq_Upar}
 \end{eqnarray}
is required to bring $U_\parallel U_h U_\Omega \vert h 
\rangle$ back into the first quadrant,  inclined at $2\theta$
above $\vert h \rangle$.  Figure~\ref{fig_coeffs} 
plots the $\vert h_\perp \rangle$ and $\vert \Omega \rangle$ 
coefficients of (\ref{eq_Uh_UOmega_gen}) as a function of 
$M/N$,  illustrating that there indeed  three regions of 
interest. 

To find the appropriate number of iterations to perform,  let
us assume $M/N \ll 1$.  Then from (\ref{eq_sin_theta_gen}) 
we will find the target state $\vert \Omega \rangle$ after 
\begin{eqnarray}
  r_* 
  \approx  
  \frac{\pi}{4\theta}
  \approx \frac{\pi}{4} \sqrt{\frac{N}{M}}
\label{eq_rNM_it}
 \end{eqnarray}
iterations.  We see that  the factor of $M$ decreases the 
number of requisite iterations relative to the factor $\sqrt{N}$ 
of the original Grover's algorithm.  This is because the search 
space becomes smaller  as $M$ increases,   and indeed,  when 
$M=N$ there is in fact no need to perform a search,  since all 
basis vectors lie in $\Omega_n = \{0,  1,  \cdots,  N-1\}$.   In fact,  
the formalism gives $\mathbb{P}_{\Omega_n}=\mathbb{1}$,  
$\mathbb{P}_\perp= 0$,   and $U_{\Omega_n}= - \mathbb{1}$.  
Since $U_{\Omega_n}$ introduces a total negative phase,  it is 
equivalent to the unit operator $\mathbb{1}$.  This formalism is 
a slight generalization of the amplitude amplification algorithm 
of Refs.~\cite{gilles}~and~\cite{grover_amp}.

\subsubsection{General Trial State}

We can further generalize the initial state of the algorithm 
from the Hadamard state to an arbitrary normalized vector 
$\vert g \rangle \equiv G \vert 0^n \rangle$,  where $G$ is 
the corresponding steering operator.  Upon decomposing 
$\vert g \rangle$ in terms of the computational basis,  
we write 
\begin{eqnarray}
  \vert g \rangle  
  &=&
  \sum_x  g_x \, \vert x \rangle
  =
  \sum_{x \in \Omega} g_x \, \vert x \rangle
  +
  \sum_{x \notin \Omega} g_x\, \vert x \rangle
  \ .
\label{eq_g_decomp}
 \end{eqnarray}
This suggest that we define the (normalized)
$\Omega$-component and orthogonal-component 
of $\vert g\rangle$ by 
\begin{eqnarray}
  \vert g_\Omega \rangle 
  &\equiv&
  {\cal N}_\Omega^{-1} \sum_{x \in \Omega} g_x \vert x \rangle
\\[5pt]
  \vert g_\perp \rangle
  &\equiv&
  {\cal N}_\perp^{-1}  \sum_{x \notin \Omega} g_x \vert x \rangle
  \ ,
 \end{eqnarray}
where the normalization factors are
\begin{eqnarray}
  {\cal N}_\Omega 
  &=&
  \left( \, \sum_{x \in \Omega} g_x \, g_x^*\right)^{1/2}
  \equiv
  \sin\theta
\label{eq_g_sintheta}
\\[5pt]
  {\cal N}_\perp 
  &=&
  \left( \, \sum_{x \notin \Omega} g_x \, g_x^*\right)^{1/2}
  =
  \cos\theta
  \ .
\label{eq_g_costheta}
 \end{eqnarray}
Furthermore,  since $\langle g_\Omega \vert g_\perp 
\rangle = 0$ and  $\langle g_\perp \vert g_\perp \rangle =
\langle g_\Omega \vert g_\Omega \rangle = 1$,  we see that
$\vert g_\Omega \rangle$ and $\vert g_\perp \rangle$
form an orthonormal basis for a 2-dimensional subspace,
which we shall call the $\Omega$-$g$ subspace.   We have
expressed the normalization factors (\ref{eq_g_sintheta})
and (\ref{eq_g_costheta}) in terms of an angle $\theta$,
and so the decomposition (\ref{eq_g_decomp}) can be
expressed as
\begin{eqnarray}
  \vert g \rangle  
  &=&
   {\cal N}_\perp \, \vert g_\perp \rangle
  +
  {\cal N}_\Omega \, \vert g_\Omega \rangle 
\\[5pt]
  &=&
   \cos\theta \, \vert g_\perp \rangle
  +
  \sin\theta\, \vert g_\Omega \rangle 
  \ .
 \end{eqnarray}
Thus,  $\theta$ is the angle of inclination of $\vert g \rangle$ 
relative to the horizontal axis $\vert g_\perp \rangle$ in the
$\Omega$-$g$ subspace.  This leads to the following Grover 
iteration.  First,  let the phase oracle $U_\Omega$ act upon the 
trial state $\vert g \rangle$,  giving 
\begin{eqnarray}
  U_\Omega \, \vert g \rangle  
  &=&
   {\cal N}_\perp \, \vert g_\perp \rangle
  -
  {\cal N}_\Omega \, \vert g_\Omega \rangle 
\\[5pt]
  &=&
   \cos\theta\,  \vert g_\perp \rangle
  -
  \sin\theta\, \vert g_\Omega \rangle 
  \ ;
 \end{eqnarray}
we then complete the iteration with the diffusion 
operation,  giving
\begin{eqnarray}
  U_g \,U_\Omega \, \vert g \rangle  
  &=&
  \Big[2 \vert g \rangle \langle g \vert  - \mathbb{1} 
  \Big]
  \Big[
   {\cal N}_\perp \, \vert g_\perp \rangle
  -
  {\cal N}_\Omega \, \vert g_\Omega \rangle 
  \Big]
\\[5pt]
  &=&
   {\cal N}_\perp \Big[ 1 - 4 \,{\cal N}_\Omega^2 \Big]\, \vert g_\perp \rangle
  +
  {\cal N}_\Omega \Big[ 3 - 4\, {\cal N}_\Omega^2 \Big] \,  \vert g_\Omega \rangle 
\\[5pt]
  &=&
   \cos3\theta \, \vert g_\perp \rangle
  +
  \sin3\theta\, \vert g_\Omega \rangle 
  \ .
 \end{eqnarray}
As before,  this vector lies in the 1-st quadrant
only when both coefficients of $\vert g_\perp \rangle$ 
and $\vert g_\Omega \rangle$ are positive (and it lies
in the 3-rd quadrant when both are negative),  otherwise 
a further Householder reflection $U_\parallel = 2 \vert 
g_\Omega \rangle \langle g_\Omega \vert - \mathbb{1}$
is required to bring it into the 1-st quadrant. 
The method reduces to the previous case when the $g_x$ 
are independent of $x$,  {\em i.e.} when $g_x = 1/\sqrt{N}$,
in which case the target vector becomes $\vert g_\Omega 
\rangle \equiv \vert \Omega \rangle$.  We now have the 
following Grover algorithm (we assume for simplicity
that both coefficients are positive):

\vskip0.4cm 
\begin{enumerate}
  \baselineskip 10pt plus 1pt minus 1pt
  \setlength{\itemsep}{3pt} 
  \setlength{\parskip}{1pt} %
  \setlength{\parsep}{0pt}  %
\item[1.] Construct the trail state $\vert g \rangle$ inclined 
at an angle $\theta$ above the $\vert g_\perp\rangle$ 
axis in the $\Omega$-$g$ subspace.  

\item[2.] Define the diffusion operator $U_g = 2 \vert g \rangle
\langle g \vert - \mathbb{1}$,  and the basis selection
operator $U_\Omega$. 

\item[3.] The state $U_\Omega \vert g \rangle$ lies in the 
4-th quadrant at an angle $\theta$ below the $g_\perp$ axis. 

\item[4.] The state $U_g U_\Omega \vert g \rangle$ lies back 
in the 1-st quadrant,  inclined at an angle $2\theta$ above 
$\vert g \rangle$, or a total angle of $3\theta$ above the
$\vert g_\perp \rangle$ axis. 

\item[5.] After $r$ iterations,  the state $(U_g U_\Omega)^r 
\vert g \rangle$ is inclined at angle $\theta_r=(2r + 1)\theta$ 
above the $\vert g_\perp \rangle$ axis.  When $\theta_r=\pi/2$ 
we will find the target state $\vert g_\Omega\rangle$ with 
unit probability.  The number of iterations is therefore $r_* = 
\pi/4\theta - 1/2$.  For $\theta \ll 1$,  then $r_* \approx C \sqrt{N/M}$.

\end{enumerate}
%

\subsubsection{Further Generalization}

Let us consider a steering set composed of arbitrary
orthonormal basis elements $\vert y \rangle$,  where
the relation with the computational basis is given by
\begin{eqnarray}
  \vert y \rangle 
  =
  \sum_x \Lambda_{yx} \, \vert x \rangle
  \ . 
 \end{eqnarray}
Since both sets of basis elements are orthonormal, 
the transformation matrix is unitary,  
\begin{eqnarray}
  \Lambda^\dagger \Lambda
  =
  \Lambda \Lambda^\dagger
  =
  \mathbb{1}
  \ . 
 \end{eqnarray}
Let us now take the steering set to consist of the $M$
elements
\begin{eqnarray}
  \Omega
  =
  \big\{ \vert y_{0} \rangle,  \vert y_{1} \rangle,  
  \cdots \vert y_{{M-1}}  \rangle \big\}
  \ ,
 \end{eqnarray}
and define the phase oracle to be
\begin{eqnarray}
   U_\Omega \, \vert y \rangle 
  &=&
  \left\{
  \begin{array}{ll}
   - | y \rangle &  {\rm for}~ y \in \Omega 
\\
  \phantom{-} | y \rangle &  {\rm for}~ y \notin \Omega 
  \end{array}
  \right. 
  ~~~\Rightarrow~~
  U_\Omega 
  =
   \mathbb{1} - 2  \sum_{y\in \Omega} 
  \vert y \rangle \langle y \vert 
  \ .
\label{eq_UOmega_y_HH_fx}
 \end{eqnarray}
Upon transforming back to the computational basis,
we can write
\begin{eqnarray}
   U_\Omega \, 
  &=&
   \mathbb{1} - 2  \sum_{x x^\prime} 
  \Omega_{x x^\prime} \, \vert x \rangle
  \langle x^\prime \vert
  \  ,
 \end{eqnarray}
where the {\em steering kernel} is given by 
\begin{eqnarray}
  \Omega_{x x^\prime} 
  =
  \sum_{y \in \Omega} \Lambda^\dagger_{x^\prime y}
  \Lambda_{y x}
  \ .
 \end{eqnarray}
Correlations between the basis elements can be encoded by 
using a non-separable kernel,  which could be used to capture 
higher order relations between the search elements.   Non-separable 
kernels lead to non-planar Grover algorithms,  with  the possibility 
for speed-up beyond quadratic.

\subsection{Circuit Implementation}
\label{sec_circuit}

We now look at the circuit implementations of the Grover 
algorithms we have discussed,  starting with the phase oracle 
$U_\omega$ that marks an arbitrary basis element $\vert 
\omega \rangle$.  As we have seen,  the phase oracle takes 
the form of the Householder reflection $U_\omega = \mathbb{1} 
- 2 \vert \omega \rangle \langle \omega \vert$.   We have also 
found it convenient to express the target state $\vert \omega 
\rangle$ by the corresponding bit string $\omega$,  and we
will continue this practice.  We shall examine the phase 
oracles in more detail for a simple simple 2-qubit system 
with computational basis states $\omega=00,  01,  10,  11$.   
To pick out the state $\omega=11$ we can employ the $CZ$ 
gate,  as this marks the target state $\vert 11 \rangle$ with 
a negative phase, 
\begin{eqnarray}
  U_{11}
  \equiv
   \mathbb{1}_2 - \vert 11 \rangle \langle 11 \vert
  =
  CZ : 
  \left\{
  \begin{array}{l}
  \vert 0 0 \rangle   \to  \phantom{-}\vert 0 0 \rangle
  \\
  \vert 0 1 \rangle   \to  \phantom{-}\vert 0 1 \rangle
  \\
  \vert 1 0 \rangle   \to  \phantom{-}\vert 1 0 \rangle
  \\
  \vert 1 1 \rangle   \to  -\vert 1 1 \rangle ~ \ .
  \end{array}
  \right.
\label{eq_cz_oracle}
 \end{eqnarray}
Note that we are using the OpenQASM convention 
in which the lowest order qubit is the control qubit.  
In a similar manner,  note that $\omega = 00$ is 
selected by the oracle
\begin{eqnarray}
  U_{00}
  \equiv   \mathbb{1}_2 - \vert 00 \rangle \langle 00 \vert
  =
  (X \otimes X) \cdot CZ  \cdot (X \otimes X) : 
  \left\{
  \begin{array}{l}
  \vert 0 0 \rangle   \to  -\vert 0 0 \rangle
  \\
  \vert 0 1 \rangle   \to  \phantom{-}\vert 0 1 \rangle
  \\
  \vert 1 0 \rangle   \to  \phantom{-}\vert 1 0 \rangle
  \\
  \vert 1 1 \rangle   \to  \phantom{-}\vert 1 1 \rangle ~ 
  \ ,
  \end{array}
  \right.
\label{eq_cz_diffuser}
 \end{eqnarray}
where $X$ is the $2\times 2$ single-qubit NOT 
operator.   Furthermore,  $\omega=01$ is marked by 
\begin{eqnarray}
  U_{01}
  &\equiv&
   \mathbb{1}_2 - \vert 01 \rangle \langle 01\vert
  =
 (X \otimes I) \cdot CZ  \cdot (X \otimes I) : 
  \left\{
  \begin{array}{l}
  \vert 0 0 \rangle   \to  \phantom{-}\vert 0 0 \rangle
  \\
  \vert 0 1 \rangle   \to  -\vert 0 1 \rangle
  \\
  \vert 1 0 \rangle   \to  \phantom{-}\vert 1 0 \rangle
  \\
  \vert 1 1 \rangle   \to  \phantom{-}\vert 1 1 \rangle
  \ ,
  \end{array}
  \right.
\label{eq_Uzeroone_def}
 \end{eqnarray}
where $I$ is the $2\times 2$ single-qubit identity operator.
In a similar manner,  we can mark the state $\omega={10}$ 
by the oracle
\begin{eqnarray}
  U_{10}
  &\equiv&
   \mathbb{1}_2 - \vert 10 \rangle \langle 10 \vert
  =
  (I \otimes X) \cdot CZ  \cdot (I \otimes X) : 
  \left\{
  \begin{array}{l}
  \vert 0 0 \rangle   \to  \phantom{-}\vert 0 0 \rangle
  \\
  \vert 0 1 \rangle   \to  \phantom{-}\vert 0 1 \rangle
  \\
  \vert 1 0 \rangle   \to  -\vert 1 0 \rangle
  \\
  \vert 1 1 \rangle   \to  \phantom{-}\vert 1 1 \rangle ~ .
  \end{array}
  \right.
\label{eq_cz_other}
 \end{eqnarray}
We can therefore express all four of the 2-qubit phase 
oracles by
\begin{eqnarray}
 && U_\omega
  =
  A_\omega \, U_{00} \, A_\omega  ~~, ~~ \text{where} ~~
  \omega \in \{00, 01,  10,  11 \} ~~\text{and}~~
\label{eq_Uomeg_Aomega_rel}
\\[5pt]
  &&A_{00} \equiv I \otimes I  ~, ~~ A_{01} \equiv I \otimes X
  ~,~~ A_{10} \equiv X \otimes I  ~,~~ A_{11} \equiv X \otimes X
  \ .
 \end{eqnarray}
Note that the steering operator $A_\omega$ is the tensor product
of $X$ gates or unit gates $I$,  where $X$ corresponds to the 1-bit 
and $I$ corresponds to the 0-bit of the bit string $\omega \in \{00, 
01, 10, 11\}$.  

\begin{figure}[h!]
\begin{minipage}[c]{1.0\linewidth}
\includegraphics[scale=0.25]{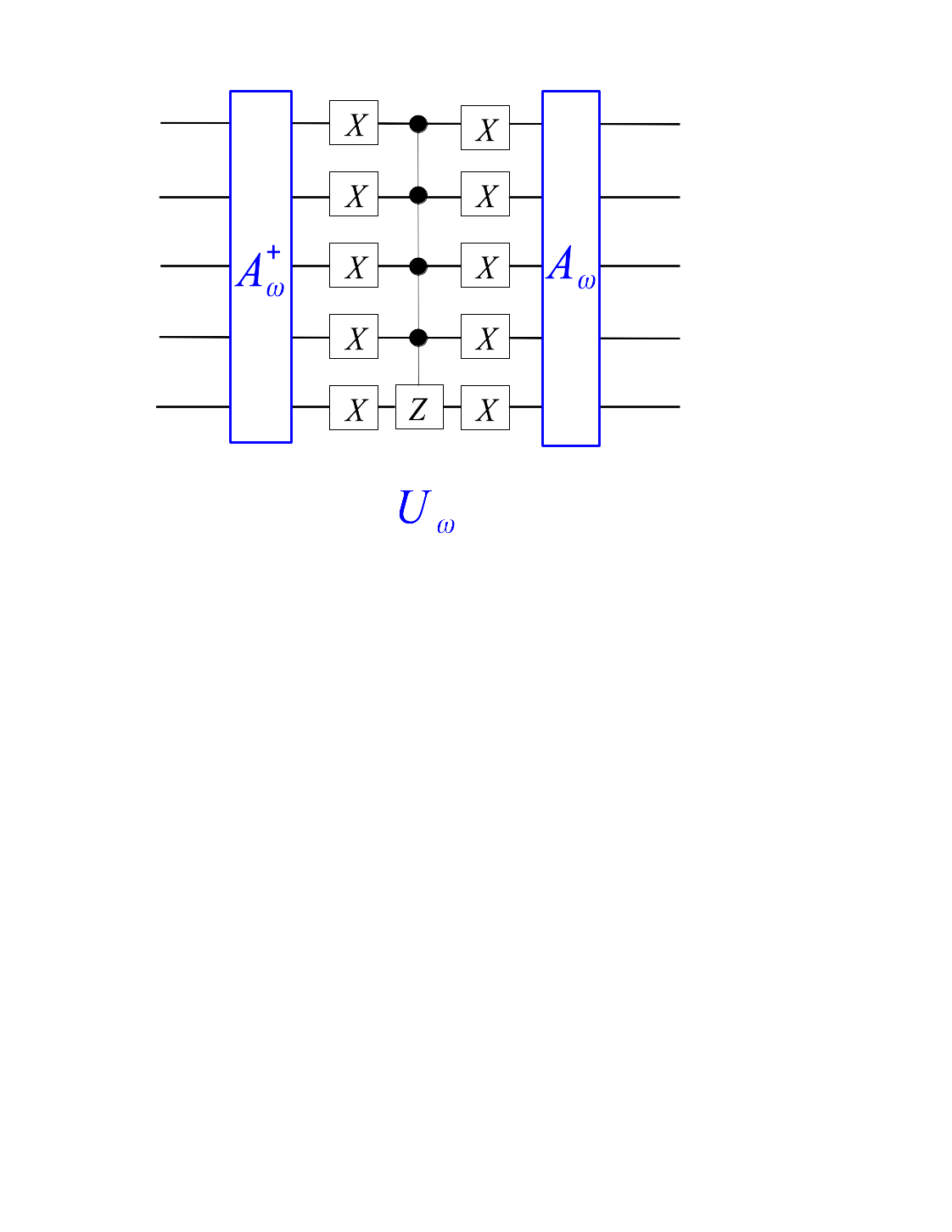} 
\vskip-7.0cm 
\caption{\footnoteskip  
Phase oracle $U_\omega = A_\omega \big[ \, \mathbb{1}_n -
2 \vert 0^n \rangle \langle 0^n \vert  \,\big] A_\omega^\dagger
= A_\omega \cdot X^{\otimes n} \cdot CC \cdots CZ \cdot
X^{\otimes n} \cdot A_\omega^\dagger$ for target bit string 
$\omega$ and state selection operator $A_\omega$.  
}
\label{fig_steering_amplitude}
\end{minipage}
\hfill
\begin{minipage}[c]{1.0\linewidth}
\includegraphics[scale=0.25]{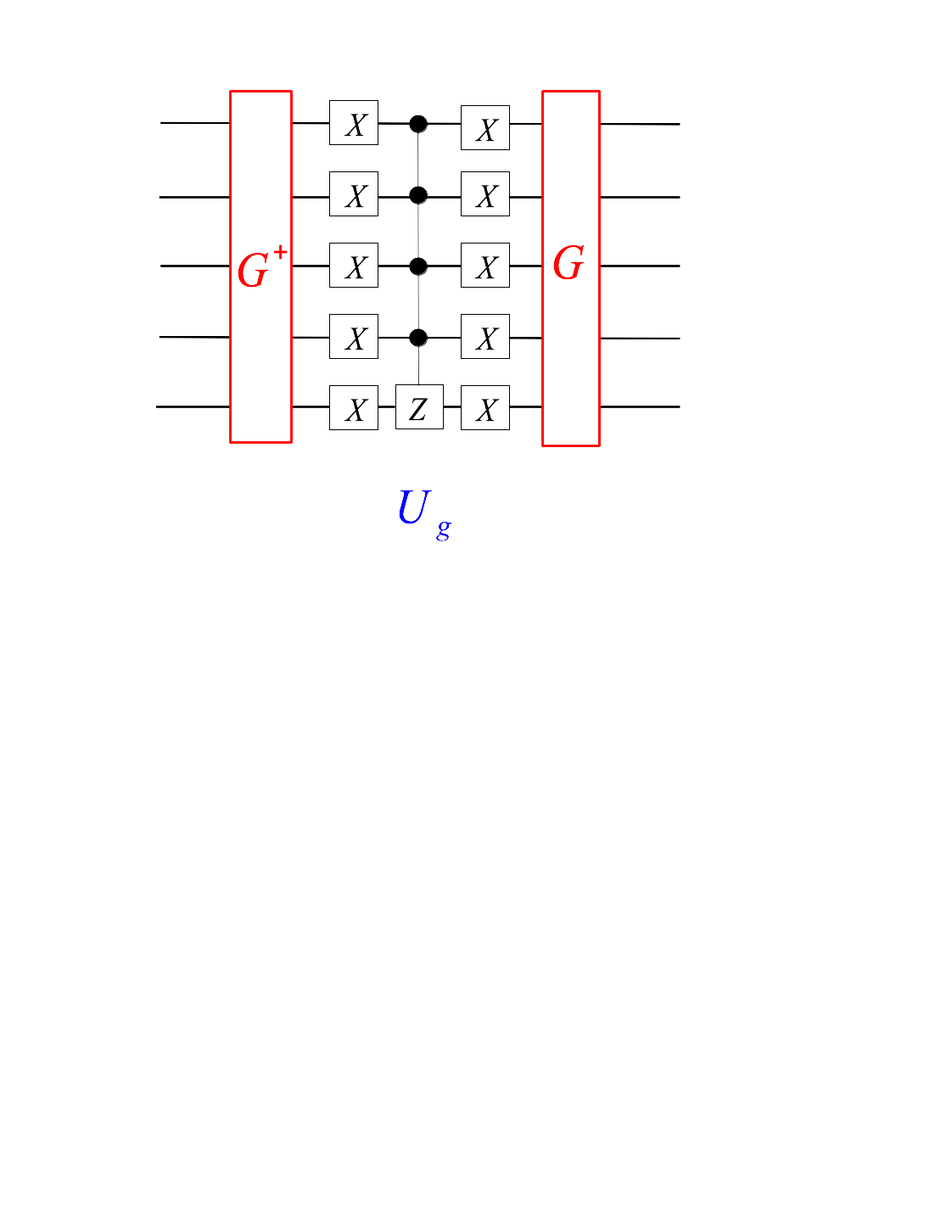} 
\vskip-7.0cm 
\caption{\footnoteskip  
  A  diffusion operator $U_g = -G \big[ \, \mathbb{1}_n - 
  2 \vert 0^n \rangle \langle 0^n \vert \,\big] 
  G^\dagger =- G \cdot X^{\otimes n} \cdot CC \cdots 
  CZ \cdot X^{\otimes n} \cdot G^\dagger$.  The quantum
  circuit is insensitive to the overall negative sign. 
}
\label{fig_grover_4qubit_diffusion_g}
\end{minipage}
\end{figure}

This construction generalizes to an arbitrary $n$-qubit system.  
For example,  equation (\ref{eq_cz_diffuser}),  which marks 
the zero-state,  becomes
\begin{eqnarray}
  U_{0 }^{(n)} 
  \equiv
  \mathbb{1}_n - 2\vert 0^n \rangle \langle 0^n \vert
  =
   X^{\otimes n} \cdot \underbrace{CC \cdots CZ}_{n-1 \,
  \text{control bits}}  \cdot \, X^{\otimes n} 
  \ .
\label{eq_Uzero_def}
 \end{eqnarray}
Upon expressing the $n$-dimensional bit string as 
$\omega = \delta_{n-1} \cdots \delta_1 \delta_0$,  where  
the individual bits are
$\delta_\ell \in \{0 , 1\}$ for $\ell \in \{0,  1,  \cdots,  n-1\}$,  
we define the state selection operators
\begin{eqnarray}
  \Delta_\ell
  =
  \left\{
  \begin{array}{ll}
  I  &:~ \delta_\ell=0
  \\
  X  &:~ \delta_\ell=1  ~~.
  \end{array}
  \right.
\label{eq_Deltaomegak_def}
 \end{eqnarray}
We then define the steering operator for state 
$\omega$ by 
\begin{eqnarray}
   A_\omega 
  \equiv
   \Delta_{n-1}   \otimes \cdots  \otimes \Delta_{1} 
  \otimes \Delta_{0}
  \ ,
\label{eq_Aomega_def}
 \end{eqnarray}
and  the $n$-qubit phase oracle becomes 
\begin{eqnarray}
  U_\omega = A_\omega \, U_{0} \, A_\omega
  \ ,
 \end{eqnarray}
where we have dropped the superscript from 
$U_0^{(n)}$ for  ease of notation.   In fact,  we can 
generalize the state preparation operator $A_\omega$
to an arbitrary unitary operator,  and define
\begin{eqnarray}
  U_\omega = A_\omega  \, U_{0} \, A_\omega^\dagger
  \ ,
\label{eq_Uomega_def}
 \end{eqnarray}
where the steering operator $A_\omega$ depends on $\omega$ 
in some functional manner,  and the requisite steering set is 
\hbox{$\Omega =\big\{ A_\omega \vert 0^n \rangle \big\}$}.  
The corresponding quantum circuit is illustrated in 
Fig.~\ref{fig_steering_amplitude}.  

\begin{figure}[h!]
\includegraphics[scale=0.25]{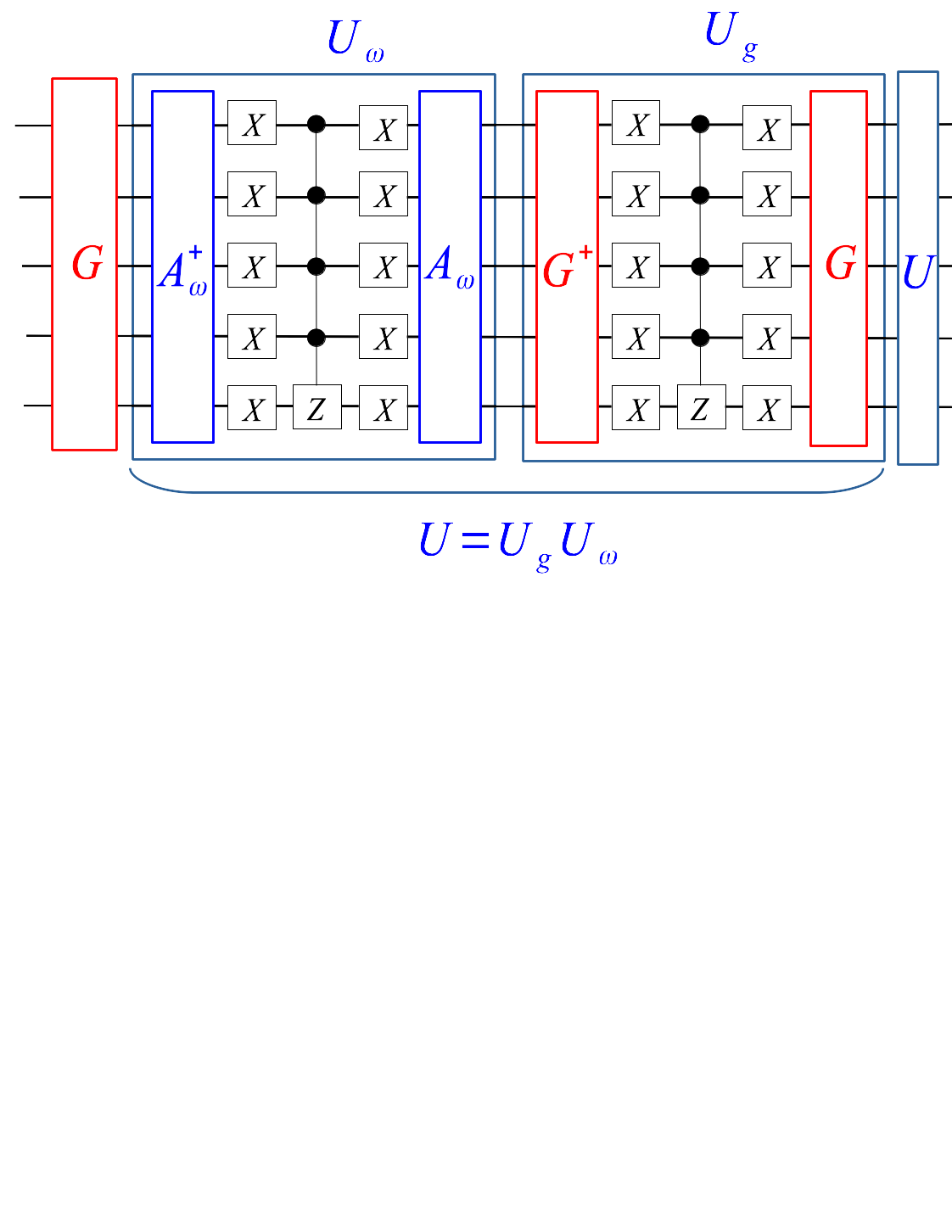} 
\vskip-7.0cm 
\caption{\footnoteskip  
  A Grover circuit for a phase oracle with steering
  operator $A_\omega$  and a diffusion oracle
  with steering operator $G$.  
}
\label{fig_steering}
\end{figure}

We now turn to the diffusion portion of the circuit,   which
performs the Householder reflection $U_h=2 \vert h \rangle 
\langle h \vert - \mathbb{1}$.   Since $H^2=I$ and $\vert h 
\rangle = H^{\otimes n} \vert 0^n \rangle$,  the diffusion 
operator becomes
\begin{eqnarray}
   U_h
   =
  - H^{\otimes n}  \Big[
  \mathbb{1} - 2\vert 0^n \rangle \langle 0^n \vert \,
  \Big] H^{\otimes n}
  \ .
 \end{eqnarray}
Note that the diffuser contains the reflection operator 
$\mathbb{1} - 2 \vert 0^n \rangle \langle 0^n \vert$,  which 
marks the $\vert 0^n\rangle$ state with a negative 
phase,  and we therefore write
\begin{eqnarray}
  U_h
  = 
  -H^{\otimes n}  \cdot  X^{\otimes n} \cdot CC \cdots CZ 
  \cdot  X^{\otimes n} \cdot  H^{\otimes n}
  \ .
 \end{eqnarray}
For steered diffusion,  in which $\vert g \rangle = G\, \vert
0^n \rangle$,  we find 
\begin{eqnarray}
  U_g
  = 
  -G  \cdot  X^{\otimes n} \cdot CC \cdots CZ  \cdot
  X^{\otimes n} \cdot  G^\dagger
  \ .
 \end{eqnarray}
This is just a generalization of the diffusion circuit in 
Ref.~\cite{3bit_grover},   and an example for $n=5$ 
qubits is illustrated in 
Fig.~\ref{fig_grover_4qubit_diffusion_g}.  Putting all 
of the pieces together gives the Grover circuit in 
Fig.~\ref{fig_steering}.
\begin{figure}[b!]
\includegraphics[scale=0.25]{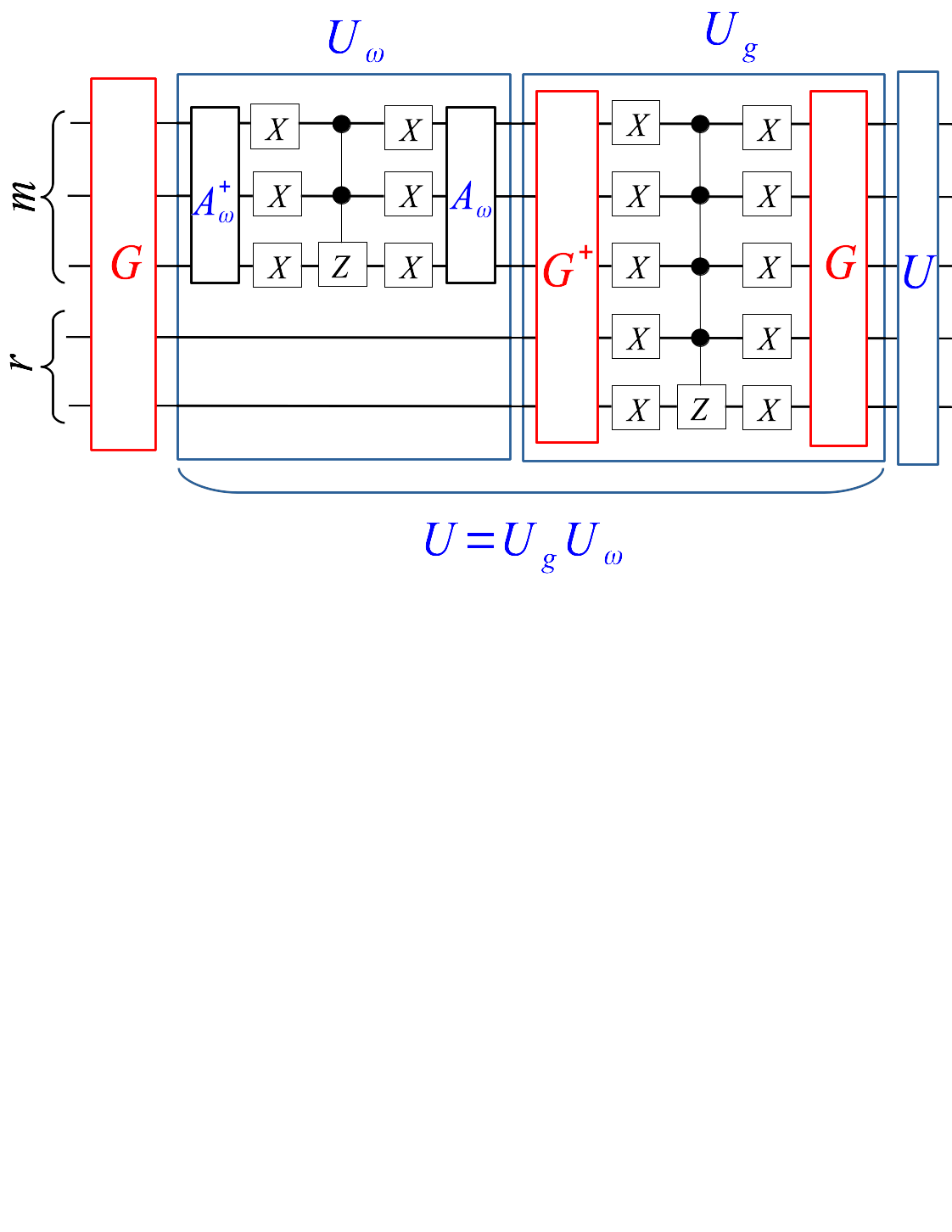} 
\vskip-7.0cm 
\caption{\footnoteskip  
  Grover circuit in which the amplitude steering operator
  is composed of an $m$-qubit subsystem of the $n$-qubit
  system.  We have dropped the index from each of the
  steering operators for ease of notation,   since their dimension
  is clear from the Figure. 
}
\label{fig_steering_diffusion_msub}
\end{figure}
Suppose now that the amplitude steering space $\mathbb{H
}_\Omega$ is an $m$-qubit subsystem with  $m < n$,  where 
the \hbox{$m$-qubits} are listed first,  {\em i.e.}  we use the
OpenQASM convention in which the first $m$-qubits 
start at bit location $2^0$ and end at bit $2^{m-1}$.    
Then the steering space has dimension $M=2^m$,  and 
the corresponding Hilbert space will be denoted by 
$\mathbb{H}_m \equiv \mathbb{H}_\Omega$.   Note 
that $\mathbb{H}_m$ is a proper subspace of $\mathbb{H}_n$,  
with orthogonal subspace $\mathbb{H}_{n-m} \equiv
\mathbb{H}_\perp$.   Letting $r \equiv n-m>0$,  we see 
that the orthogonal space has dimension $R=2^{n-m} = 
2^r$,  and from our qubit ordering convention we have 
$\mathbb{H}_n = \mathbb{H}_r \otimes \mathbb{H}_m$.  
Recall that the unit operator has the decomposition
$\mathbb{1}_n = \mathbb{1}_r \otimes \mathbb{1}_m$,
and that $\Omega_n$ ,  $\Omega_r$ and $\Omega_m$ 
denote the respective computational basis sets for 
$\mathbb{H}_n$,  $\mathbb{H}_r$ and $\mathbb{H}_m$.  
Let $A_\omega^m$ be a unitary steering operator on 
$\mathbb{H}_m$,  {\em i.e.} $A_\omega^m \, A_\omega^{
m\, \dagger} = \mathbb{1}_m$.  Suppose we wish 
to mark the state $A_\omega^m \vert 0^m \rangle \in 
\mathbb{H}_m$ in the $m$-qubit system (without regard 
to the $r$-qubit system).  The appropriate steering set 
is therefore 
\begin{eqnarray}
  \Omega_\omega
  = 
  \Big\{ \vert y \rangle_n \equiv
  \vert x \rangle_r \otimes A_\omega^m\, \vert 0^m \rangle
  \, \Big\vert \, x \in \Omega_r \Big\}
  \ ,
\end{eqnarray}
and the phase oracle can be expressed as (upon
replacing $U_{\Omega_\omega}$
by the simpler notation $U_\omega$): 
\begin{eqnarray}
  U_\omega
  &\equiv&
   \mathbb{1}_n - 2  \sum_{y\in \Omega_\omega} 
  \vert y \rangle \langle y \vert_n
   =
  \mathbb{1}_r \otimes \mathbb{1}_m
  - 2  
  \underbrace{\sum_{x\in \Omega_r} \vert x \rangle 
  \langle x \vert_r}_{  \mathbb{1}_r  }
  \otimes 
  A_\omega^m \vert 0^m \rangle \langle 0^m \vert\, 
  A_\omega^{m\,\dagger}
\\[5pt]
  &=&
  \mathbb{1}_r \otimes
  A_\omega^m \, 
  \Big[  \mathbb{1}_m  - 2 \vert 0^m \rangle
  \langle 0^m \vert \, \Big] \, A_\omega^{m \, \dagger}
  \ .
\label{eq_Uomega_reduced_m}
 \end{eqnarray}
In factoring out the unit matrix $\mathbb{1}_r$ for $\mathbb{H}_r$,  
we have used the completeness relation $\sum_{x \in 
\Omega_r} \vert x \rangle \langle x \vert_r = \mathbb{1}_r$.
We consequently find the quantum circuit illustrated in 
Fig.~\ref{fig_steering_diffusion_msub}.  As we shall see 
in the next section,  this is precisely the circuit developed 
by Hiroyuki {\em et al.}.  

\clearpage
\section{Quantum Pattern Matching}
\label{sec_pattern}

Hiroyuki {\em et al.}\,\cite{image_grover}  have recently provided 
a working version of Grover's algorithm that  implements an 
unstructured database search for image pattern matching. 
Recall that there have been two primary difficulties in creating 
a practical working Grover search: the dictionary problem 
and the exponential gate problem.   Hiroyuki {\em et al.} solve 
the dictionary problem by mapping the classical database 
entries into a well chosen quantum multi-qubit system whereby
the database entries correspond basic pixel formations.  
More specifically,  they use a  4-pixel binary data structure 
in which each quadrant can be black or white (on or off),   
thereby giving $2^4=16$ possible pixel choices that form 
the words of the dictionary.  These 16 base structures are 
then mapped onto the computational bases of 4-qubit 
subsystems.  Even more 
significantly,  they solve the exponential gate problem by 
employing  a new method called {\em approximate amplitude 
encoding} (AAE)\,\cite{aae21}.  This method approximates 
the exponentially large database by a constant depth  {\em 
parameterized quantum circuit} containing a polynomial 
number of gates,  and the parameters of the circuit are 
trained using a machine learning technique.  In this section,  
we derive the quantum circuit used by Hiroyuki {\em et al.} 
as a special case of the steering formalism.  Since we use 
OpenQASM conventions and Hiroyuki {\em et al.}  employ the 
standard physics conventions of qubit ordering,  one must 
adjust the quantum circuit for the difference in conventions.

\subsection{Mathematical Statement of the Problem}

Before presenting the Grover algorithm,  we formulate the 
search problem in a precise mathematical fashion.   Suppose 
we have a {\em database} composed of $R$ complex valued 
data vectors ${\bf a}_k$,   each of dimension of $M$,  
\begin{eqnarray}
  {\bf a}_k 
  = 
  (a_{0 \,k},  a_{1 \,k},  \cdots,  a_{{M-1}\, k})^\smT
  ~~\text{with}~~ k \in \{0, 1, \cdots,  R-1  \}
  \ ,
\end{eqnarray}
where the elements $a_{j k}$ are complex numbers that 
define the entries in the database  for $j \in \{ 0,  1,  \cdots,  
M-1\}$.   Let us further suppose that we are  
given an $M$-dimensional {\em query vector} of
complex numbers
\begin{eqnarray}
  {\bf b}
  = 
  (b_{0},  b_{1},  \cdots,  b_{{M-1},})^\smT
  \ .
\end{eqnarray}
The query vector represents the target state that we wish to 
find in the database.  Since the vectors ${\bf a}_k$ and ${\bf b}$ 
are $M \times 1$ column vectors,  we have expressed them by 
the transpose of $1 \times M$ row vectors (so that the
expressions will fit on a single horizontal line of text).   Our 
goal is to identify the index $k=k_*$ for which the query
vector ${\bf b}$ has the largest overlap with the data vector
${\bf a}_k$,  
\begin{eqnarray}
  k_* = \text{argmax}_k \vert {\bf b}^\dagger {\bf a}_k \,    \vert
  \ .
\end{eqnarray}
Hiroyuki {\em et al.} take the database vectors and the query 
vector 
to have real entries,  but we have generalized them to complex 
values.   The practical reason for restricting $a_{jk}$ and $b_j$ 
to be real is because AAE is only applicable to real-valued 
circuits.  Nonetheless,  we feel there is value in generalizing the 
problem to complex vectors,  in which case,  AAE must be 
applied separately to the real an imaginary parts of the steering 
operator (or a more general approximate amplitude encoding 
method must be devised).

If we are given a black-box that calculates $\vert {\bf b}^\dagger 
{\bf a}_k\vert$,  then a  classical algorithm would require $R$ 
calls to find the maximal index $k_*$.   This is because we 
would have to compare $\vert {\bf b}^\dagger {\bf a}_k \,    
\vert$ for all $R$ possible values of $k$,  and then select 
the index $k=k_*$ for which $\vert {\bf b}^\dagger {\bf a}_k \,   
\vert$ has the largest value.  Note that 
we have defined the problem in terms of the inner product
${\bf b}^\dagger {\bf a}_k$ (rather than using the opposite
but equivalent ordering ${\bf a}_k^\dagger\,  {\bf b})$ to remind 
ourselves that the quantum counterpart 
of this classical problem involves the {\em transition amplitude} 
${\cal A}$ from an {\em initial} quantum state $\vert {\bf a_k} 
\rangle$ determined by the database to a {\em final} quantum 
state $\vert {\bf b} \rangle$ determined by the query vector,   
so that $ {\cal A} = \langle {\bf b}\vert {\bf a}_k \rangle$.   
Grover's algorithm maximizes the overlap between the 
states $\vert{\bf a}_k \rangle$ and $\vert {\bf b} \rangle$, 
finding the correct index $k_*$ in approximately $\sqrt{R}$ 
calls to the database. 

\subsection{Encoding the Database in a Quantum State}

Let us now consider two quantum systems with $m$ and $r$ 
qubits,  so that the number of computational basis states in 
each system is $M = 2^{m}$ and $R = 2^r$.  The $m$-system 
is used to store the data entries ${\bf a}_k$,  and the $r$-system 
stores the data index $k$.  We order the $m$-qubit data 
system first (the lower order bits),  followed by the $r$-qubit 
index system,  and we denote the corresponding Hilbert spaces 
by $\mathbb{H}_m$ and $\mathbb{H}_r$.  We also denote the
respective computational basis states for the data and
index spaces by
\begin{eqnarray}
  \vert j \rangle_m ~~\text{and}~~ \vert k \rangle_r
  \ ,
\end{eqnarray}
where the state indices range over
\begin{eqnarray}
  j \in \Omega_m \equiv \{0,  1,  \cdots,  M-1\} 
  ~~\text{and} ~~k\in \Omega_r \equiv \{0,  1,  \cdots,  R-1\}
  \ .
\end{eqnarray}
Recall that $\Omega_m$ and $\Omega_r$ are the set of all 
computational basis elements for $\mathbb{H}_m$ and
$\mathbb{H}_r$,  respectively.  
For clarity we have indicated the qubit space to which the
state belongs by an $m$- or an \hbox{$r$-subscript},  keeping
in mind that the index $j$ is associated with the database
entry,  and the index $k$ corresponds to the database index.
Let us  now define quantum states in $\mathbb{H}_m$ that 
correspond to the $k$-th database entry ${\bf a}_k$ and 
to the query vector ${\bf b}$  as follows,  
\begin{eqnarray}
  \vert \text{data}(k)\rangle_m
  &\equiv&
  \sum_{j \in \Omega_m} a_{j k } \, \vert j \rangle_m
  \in \mathbb{H}_m
  ~~~\text{for}~~ k \in \Omega_r
\\[5pt]
  \vert \text{query} \rangle_m
  &\equiv&
   \sum_{j \in \Omega_m} b_j \, \vert j \rangle_m
  \in \mathbb{H}_m
  \ .
\label{eq_query_m_def_one}
\end{eqnarray}
Since we have encoded the  database and query 
vectors into quantum states,  we assume that 
\begin{eqnarray}
  \sum_{j \in \Omega_m}  \vert a_{jk} \vert^2 =1 
  ~~\forall k \in \Omega_r
  ~~~\text{and}~~~
  \sum_{j \in \Omega_m} \vert b_j\vert^2 =1
  \ .
\end{eqnarray}
To finish the quantum database construction,  we form a 
composite qubit system from the $m$-qubit data space 
and the $r$-qubit index space.    Taking \hbox{$n = r+m$},  
we denote this $n$-qubit database system by $\mathbb{H}_n 
= \mathbb{H}_r \otimes \mathbb{H}_m$,  and note that 
the computational basis elements are given by  
\begin{eqnarray}
  \vert x \rangle_n
  \equiv
  \vert k \rangle_r \otimes \vert j \rangle_m
  \in \mathbb{H}_n
  \ .
\label{eq_yxz_basis}
 \end{eqnarray}
To form a database 
entry in $\mathbb{H}_n$,  we concatenate the  state $\vert 
\text{data}(k)\rangle_m \in \mathbb{H}_m$ with the index 
state $\vert k \rangle_r \in \mathbb{H}_r$  to form 
 \hbox{$\vert k \rangle_r \otimes \vert 
\text{data}(k)\rangle_m \in \mathbb{H}_n$}.  Upon taking 
a uniform sum over $k$,  we arrive at the final form of 
the quantum database,  
\begin{eqnarray}
  \vert \text{database} \rangle_n
  &\equiv&
  A_n \, \vert 0 \rangle^{\otimes n }
\\[5pt]
  &\equiv&
  \frac{1}{\sqrt{R}} \sum_{k \in \Omega_r}
  \vert k \rangle_r \otimes \vert \text{data}(k)\rangle_m
\\[5pt]
  &=&
  \frac{1}{\sqrt{R}}
  \sum_{j \in \Omega_m}  \sum_{k \in \Omega_r}
  a_{j k} \,\vert k \rangle_r \otimes  \vert j \rangle_m 
  \ ,
\label{eq_database_query_A}
\end{eqnarray}
where we have introduced the unitary steering operator 
$A_n$ that generate the database state from the corresponding 
zero-state.
We also introduce a steering operator $B_m$ for 
the query state,  reexpressing (\ref{eq_query_m_def_one}) as
\begin{eqnarray}
  \vert \text{query} \rangle_m
  &\equiv&
  B_m \, \vert 0 \rangle^{\otimes m}
\\[5pt]
  &\equiv&
 \sum_{j \in \Omega_m}   b_j \, \vert j \rangle_m 
  \ .
\label{eq_database_query_B}
\end{eqnarray}
We have used $n$- and $r$-subscripts on the steering 
operators to emphasize the Hilbert spaces on which these
operators act,  
%
\begin{eqnarray}
    A_n &:&  \mathbb{H}_n \to \mathbb{H}_n
\\
    B_m &:&  \mathbb{H}_m \to \mathbb{H}_m
  \ .
\end{eqnarray}
Unless we wish to stress the dimension of the qubit 
system,  we shall drop the qubit subscripts from $A_n$ 
and $B_m$,  and simply write $A$ and $B$.  The overlap 
in the $m$-qubit space between the query state and a 
data entry is
\begin{eqnarray}
  {\cal A}_m(k)
  \equiv
  \langle \text{query}  \vert \text{data}(k) \rangle_m
  &=&
  \langle 0^m \vert
  \, B^\dagger \vert \text{data}(k) \rangle_m
  \ .
\label{eq_Am_def}
\end{eqnarray}
This suggests that our amplitude steering set $\Omega$ 
should be chosen to mark the zero-state $\vert 0^m \rangle 
\equiv \vert 0 \rangle^{\otimes m}$ of the $m$-system
(without regard to the $r$-system). 
However,  the state  $\vert \text{data}(k)\rangle_m$ in not 
directly accessible to us.  Instead,  we only know the total 
database state
\begin{eqnarray}
  \vert \text{database}\rangle_n 
  = 
  A\, \vert 0^n \rangle
  \in \mathbb{H}_n 
  \ .
\end{eqnarray}
We also know the $m$-qubit query state $\vert\text{query} 
\rangle_m \in \mathbb{H}_m$.  To properly compare these states, 
we define the enlarged query state
\begin{eqnarray}
  \vert \overline{\text{query}} \rangle_n
  &\equiv&
  (\mathbb{1}_r \otimes B) \vert 0^n \rangle
  \in \mathbb{H}_n 
  \ .
\end{eqnarray}
We shall express the unit operators on $\mathbb{H}_n$,  
$\mathbb{H}_m$ and $\mathbb{H}_r$ by $\mathbb{1}_n$,  
$\mathbb{1}_m$ and $\mathbb{1}_r$,  respectively,  and 
because of our qubit ordering,  we have $\mathbb{1}_n 
= \mathbb{1}_r  \otimes \mathbb{1}_m$.  Upon restoring 
the qubit indices on the steering operators for clarity,   the 
overlap between the database and the enlarged query 
state is given by 
\begin{eqnarray}
  {\cal A}_n
  \equiv
  \langle\overline{\text{query}} \vert
  \text{database} \rangle_n
  =
  \langle 0^n \vert
  (\mathbb{1}_r \otimes B_m^\dagger) A_n \, \vert 0^n 
  \rangle
  \ .
\end{eqnarray}
This suggests that we select the trial state for the first
Grover iteration to be 
\begin{eqnarray}
  \vert g \rangle
  &=&
  \big( {\mathbb{1}}_r  \otimes B_m^\dagger \big) A_n \, 
  \vert 0^n \rangle
  \equiv
  G_n  \,  \vert 0^n \rangle
  \ ,
\end{eqnarray}
where $G_n$ is the steering operator for the $n$-qubit
system.   Thus,  the diffusion operator takes the form 
\begin{eqnarray}
  U_g 
  &=& 
  2 \vert g \rangle \langle g \vert - \mathbb{1}_n
\\[5pt]
    &=&
  - G\, \Big[  \, \mathbb{1}_n - 2 \vert 0^n \rangle
  \langle 0^n \vert  \, \Big] G^\dagger
  \ .
\end{eqnarray}
\begin{figure}[t!]
\includegraphics[scale=0.25]{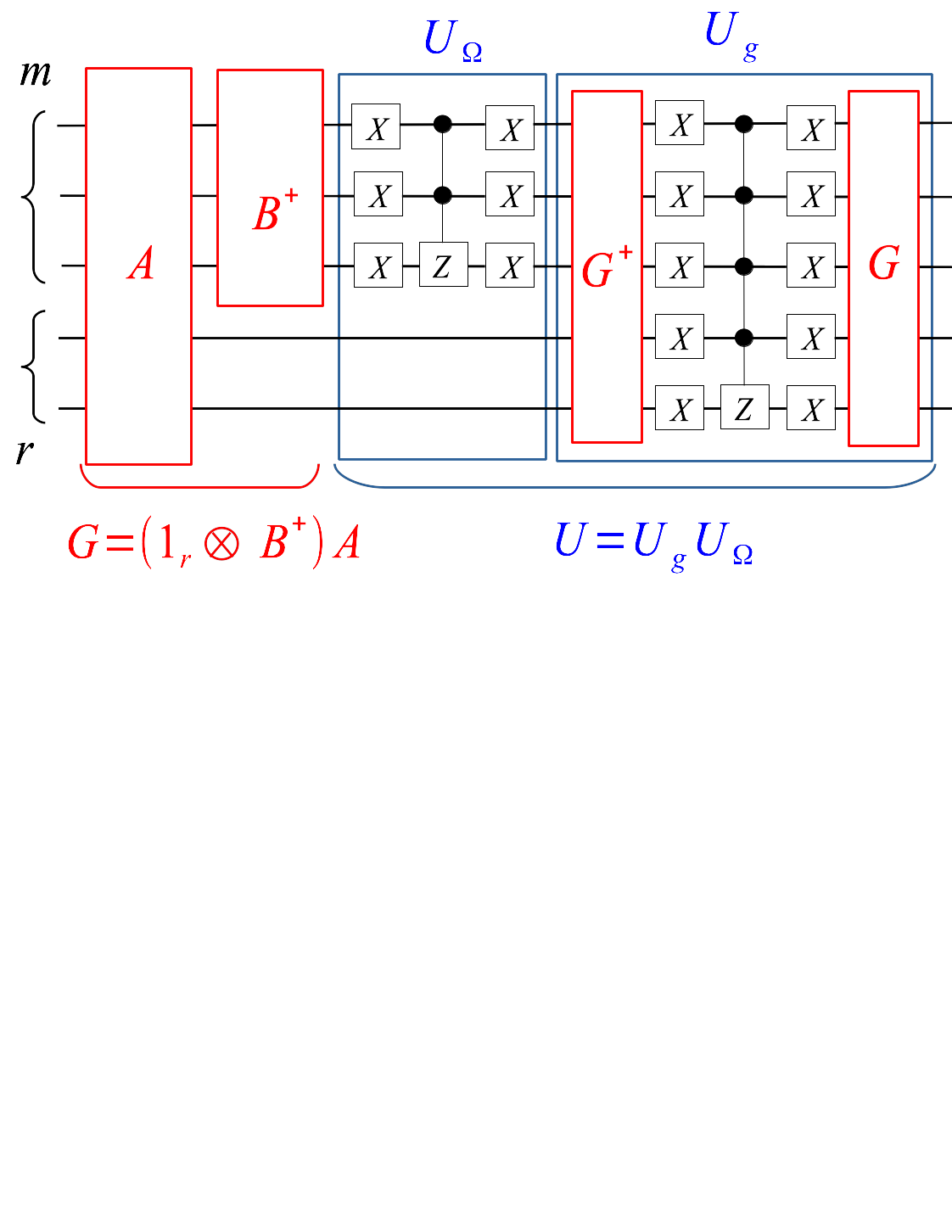} 
\vskip-6.5cm 
\caption{\footnoteskip  
  The Grover circuit of Hiroyuki {\em et al.} for quantum
  database search.  The database state is encoded in
  the operator $A_n$,  and the query state is encoded by
  the operator $B_m$.  The amplitude oracle is $U_\Omega 
  = \mathbb{1}_r \otimes \big[ \mathbb{1}_m - 2 \vert 0^m 
  \rangle  \langle 0^m \vert \, \big]$.   The trial 
  state is taken to be $\vert g \rangle = G_n \vert 0^n \rangle$,
  where $G_n = \big(\mathbb{1}_r \otimes B_m^\dagger \big) 
  A_n$,  and the diffusion operator  is $U_g = -G_n  \big[ \, 
  \mathbb{1}_n - 2 \vert 0^n \rangle\langle 0^n \vert   \, \big] \, 
  G_n^\dagger$.   
  In the Figure,  we have dropped  the subscripts on $A$,  $B$,  
  and $G$ to avoid clutter. 
}
\label{fig_tezuka}
\end{figure}
As we have noted,  the amplitude (\ref{eq_Am_def}) 
suggests that we mark the $m$-qubit zero state 
$\vert 0 \rangle_m$,   and so we define the amplitude 
steering set to be 
\begin{eqnarray}
  \Omega 
  = 
  \Big\{ \vert k \rangle_r \otimes\vert 0^m \rangle\,
  \Big\vert\, k \in \Omega_r \Big\}
  \ .
\end{eqnarray}
This steering set marks the zero-state in the \hbox{
$m$-system},  and any index element in the \hbox{
$r$-system}.  From definition (\ref{eq_UOmega_HH_fx}) 
we see that the phase oracle takes the form 
\begin{eqnarray}
  U_{\Omega}
  &=&
   \mathbb{1}_n - 2  \sum_{x\in \Omega} 
  \vert x \rangle \langle x \vert_n
   =
  \mathbb{1}_n
  - 
  2  \sum_{k\in \Omega_r} \vert k \rangle \langle k \vert_r
  \otimes 
  \vert 0^m \rangle \langle 0^m \vert
\\[5pt]
  &=&
  \mathbb{1}_n
  - 
  2  \, \mathbb{1}_r
  \otimes
  \vert 0^m \rangle\langle 0^m \vert
  \ ,
 \end{eqnarray}
where we have used the completeness relation $\sum_{k \in 
\Omega_r} \vert k \rangle \langle k \vert_r = \mathbb{1}_r$. 
Upon decomposing \hbox{$\mathbb{1}_n = \mathbb{1}_r 
\otimes \mathbb{1}_m$},   we therefore find 
\begin{eqnarray}
  U_\Omega
  =
  \mathbb{1}_r \otimes \Big[
  \mathbb{1}_m - 2 \vert 0^m \rangle \langle 0^m \vert \,
  \Big]
  \ .
\end{eqnarray}
Figure~\ref{fig_tezuka} represent the corresponding quantum 
circuit,  which is the same as the circuit presented in Hiroyuki 
{\em et al.} (adjusted for OpenQASM conventions).

We can recast this circuit in a more intuitive manner.  Upon
replacing the qubit subscripts on the steering operator
for clarity,  let us
select a steering set that marks the query state $\vert 
\text{query} \rangle_m = B_m\,\vert 0^m\rangle \in
\mathbb{H}_m$,  namely
\begin{eqnarray}
  \Omega 
   =
  \Big\{ \vert k \rangle_r \otimes 
  B_m\, \vert 0^m \, \rangle\,  \Big\vert\, k \in \Omega_r \Big\}
  \ ,
\label{eq_Omega_B_def}
\end{eqnarray}
where $\vert k \rangle_r \in \Omega_r$ is any computational 
basis set of the $r$-qubit index system.  We also choose the 
initial state to be steered by the database operator $A_n$,  
\begin{eqnarray}
  \vert g \rangle
  =
  A_n \,  \vert 0^n \rangle
  \in \mathbb{H}_n 
  \ .
\label{eq_Omega_A_def}
\end{eqnarray}
\begin{figure}[t!]
\includegraphics[scale=0.25]{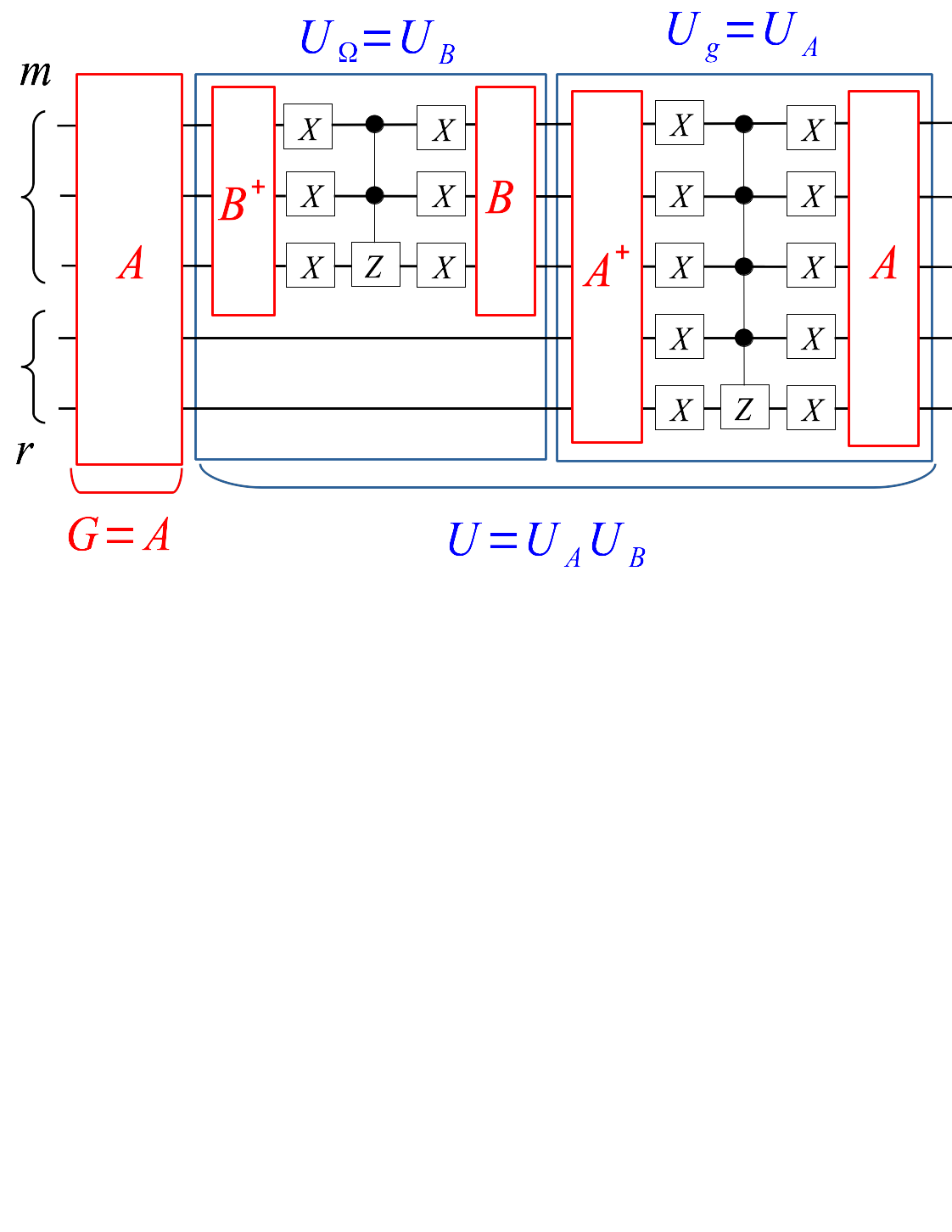} 
\vskip-6.5cm 
\caption{\footnoteskip  
  A reorganized Grover circuit for quantum database search, where
  the amplitude oracle \hbox{$U_\Omega = \mathbb{1}_r \otimes 
  B_m  \,\big[ \mathbb{1}_m - 2 \vert 0^m \rangle  \langle 0^m \vert 
  \, \big] B_m^\dagger \equiv U_\smB$} is steered by the query 
  operator $B_m$.  The trial  state $\vert g \rangle = A_n \vert 0^n 
  \rangle$ is steered by the database operator $A_n$,  and the 
  diffusion operator is therefore \hbox{$U_g = -A_n \, \big[ \,\mathbb{1}_n 
  - 2 \vert 0^n  \rangle \langle 0^n \vert \, \big]\, A_n^\dagger \equiv
  U_\smA$}.    For ease of notation,   we have dropped the qubit indices 
  from the operators in the Figure.  
}
\label{fig_tezuka_two}
\end{figure}
The steering set (\ref{eq_Omega_B_def}) now gives the phase oracle
\begin{eqnarray}
  U_\Omega
  =
  \mathbb{1}_r \otimes B_m \Big[
  \mathbb{1}_m - 2 \vert 0^m \rangle \langle 0^m \vert \,
  \Big]  B_m^\dagger
  \ ,
\end{eqnarray}
which corresponds to the oracle presented in 
(\ref{eq_Uomega_reduced_m}).  Furthermore,  
(\ref{eq_Omega_A_def}) implies that the diffusion 
operator is now given by 
\begin{eqnarray}
  U_g 
  &=& 
  2 \vert g \rangle \langle g \vert - \mathbb{1}_n
\\[5pt]
    &=&
  - A_n\, \Big[  \, \mathbb{1}_n - 2 \vert 0^n \rangle
  \langle 0^n \vert  \, \Big] A_n^\dagger
  \ .
\end{eqnarray}
The complete Grover circuit for these choices is illustrated 
in   Fig.~\ref{fig_tezuka_two}.  Note that this is analogous 
to the circuit of Fig.~\ref{fig_steering_diffusion_msub},
and that the database operator $A_n$ fixes the initial state 
$\vert g \rangle$ (along with the diffusion operator),  while 
the query operator $B_m$ determines the phase oracle 
$U_\Omega$.  When applying AAE,  however,  it is more 
convenient to associate $A_n$ and $B_m$ together into a 
single operator $G_n=\big(\mathbb{1}_r \otimes 
B_m^\dagger \big) A_n$,  as Hiroyuki {\em et al.} do. 
However,  the second approach makes the rolls of $A_n$
and $B_m$ physically clear.  

\clearpage
\section{Conclusions}
\label{sec_conclusions}

This paper provides an overview of Grover's original algorithm 
using well motivated physical arguments.   Grover's algorithm 
starts by choosing a trial wave function as a first guess.  This 
state is then acted upon by a phase oracle that marks the wave 
function in the direction of the target state.   The resulting 
state is then sent through a diffusion operation to enhance 
the marked component,  and the process is repeated until 
the target state is achieved with unit probability.   This paper 
introduces the notion of steering operators meant to bias 
the diffusion and amplitude selection processes.  We provide 
a number of generalizations to Grover's algorithm using the
steering operator formalism,  recasting amplitude amplification 
in terms of these operators.   In particular,  our formalism 
accommodates an arbitrary target set,  rather than a target 
state consisting of a single computational basis element.  
We also generalize the amplitude oracle to capture higher 
order correlations that might exist between the dictionary
elements.  This is performed by introducing a non-separable 
kernel into the oracle,  which leads to non-planar Grover 
algorithms  with the potential for speed-up beyond quadratic. 
We construct several quantum circuits that implement 
these generalized algorithms.  Finally,  we use the steering 
formalism to derive the quantum pattern matching 
circuit of  Hiroyuki {\em et al.}  as a special case.

%
\clearpage
\pagebreak

\end{document}